\documentclass{aa}
\usepackage{txfonts}
\usepackage{graphicx}
\usepackage{natbib}
\bibpunct{(}{)}{;}{a}{}{,} % to follow the A&A style
\begin{document}

\title{Classifying radio emitters from the Sloan Digital Sky Survey}
\subtitle{Spectroscopy and diagnostics}

\author{M. Vitale\inst{1,2}
  \and J. Zuther\inst{1}
  \and M. Garc\'{i}a-Mar\'{i}n\inst{1}
  \and A. Eckart\inst{1,2}
  \and M. Bremer\inst{1}
  \and M. Valencia-S.\inst{1,2}
  \and A. Zensus\inst{2}}
   
%\thanks{\emph{Present address: I. Physikalisches Institut, Universit\"at zu K\"oln, Z\"ulpicher Strasse 77, 50937 K\"oln, Germany}}

\offprints{Mariangela Vitale, \email{vitale@ph1.uni-koeln.de}}

\institute{I. Physikalisches Institut, Universit\"at zu K\"oln, Z\"ulpicher Strasse 77, 50937 K\"oln, Germany
  %\and International Max Planck Research School for Astronomy and Astrophysics, Auf dem H\"ugel 69, 53121 Bonn, Germany
  \and Max-Planck Instutut f\"ur Radioastronomie, Auf dem H\"ugel 69, 53121 Bonn, Germany}

\abstract {The Sloan Digital Sky Survey (SDSS) allows us to classify galaxies using optical low-ionization emission-line diagnostic diagrams. A cross-correlation of the SDSS data release 7 (DR7), containing spectroscopic data, with the Very Large Array (VLA) survey Faint Images of the Radio Sky at Twenty-centimeters (FIRST), makes it possible to conduct a joined multiwavelength statistical study of radio-optical galaxy properties on a very large number of sources.}{Our goal is to improve the study of the combined radio-optical data by investigating whether there is a correlation between the radio luminosity at $20$ cm over the luminosity of the optical H$\alpha$ line ({$L_{20~cm}/L_{H\alpha}$) and line excitation ratios, where the latter provide the spectroscopic classification in Seyferts, low-ionization nuclear emission-line regions (LINERs), and star-forming galaxies. We search for a trend with $z$ in the classification provided by classical and more recent optical emission-line diagnostic diagrams}.}{We cross-matched the optical sources with the FIRST radio survey in order to obtain spectroscopic information of a selected sample of radio emitters with optical observed counterpart. We searched for an $L_{20~cm}/L_{H\alpha}$ threshold value above which the radio emitters start being classified as active galactic nuclei (AGNs) rather than star-forming galaxies (SFGs). We investigated the origin of emission-lines by using both photoionization and shock models.}{The percentage of detected AGNs (Seyferts and LINERs) or composites is much higher in the optical-radio sample than in the optical sample alone. We find a progressive shift in the sources towards the AGN region of the diagram with increasing $L_{20~cm}/L_{H\alpha}$, with an indication of different behavior for LINERs and Seyferts. The classification appears to slightly depend on the redshift. The diagnostic diagrams display a density peak in the star-forming or composite region for log$(L_{20~cm}/L_{H\alpha})<0.716$, while the distribution in the LINER region peaks above this threshold. A comparison with photoionization and shock models shows that the large fraction of LINERs identified in our study have emission lines that may be explained by shocks.}{Our results indicate that it is worthwhile to further explore the radio domain, probing the physical nature of LINERs, thanks to a combination of optical and radio information. The {[N\sc ii]}/H$\alpha$ vs. equivalent width of the H$\alpha$ line (WHAN) diagram confirms the LINER classification for most of those that have been identified with the traditional diagnostic diagrams. The correlation between $L_{20~cm}/L_{H\alpha}$ and optical emission line ratios suggests the nuclear origin of the emission from the most powerful radio galaxies.}

\maketitle 

%\authorrunning

%\bibpunct{(}{)}{;}{a}{}{,}

\section{Introduction}
% radio context
Our knowledge of global galaxy properties has been recently improved thanks to large-area surveys, such as the Sloan Digital Sky Survey \citep[SDSS;][]{York2000}, which have allowed for statistical studies on large samples of galaxies. SDSS data can be used as a starting point to conduct multiwavelength studies on galaxy properties outside the optical regime, such as the radio domain, where active galactic nuclei (AGNs) can be detected.\newline
Radio galaxies are active galaxies that are very luminous at radio wavelengths. They are considered as radio-loud if they have ratios of radio (at $5$ GHz) to optical (B-band) flux greater than ten \citep{Kellermann1989}. According to \citet{Fanaroff1974}, radio-loud galaxies can be divided into two major classes. In Fanaroff-Riley Class I (FR I) sources, the radio emission peaks near the galaxy nucleus and the emission from the jets fades with distance from the center, while class II (FR II) sources present bright radio lobes. Class I sources dominate the population of radio emitters at low radio power and low redshifts, while more powerful radio galaxies %(with $178$-MHz radio powers greater than $10^{27}$ WHz$^{-1}$)
are almost exclusively FR II systems that can be detected at higher redshift.\newline
Radio galaxies can also be classified according to whether they have strong high-excitation narrow-line emission. The majority of FR I radio galaxies show very weak or completely absent emission lines. Those are referred to as low-excitation systems \citep{Hardcastle2006}, and they are mostly found in elliptical galaxies with little ongoing star formation \citep{Ledlow1995,Govoni2000}. If optical spectroscopic information is available, these galaxies are generally classified as low-ionization nuclear emission-line regions \citep[LINERs,][]{Heckman1980}. Conversely, the most powerful, high-redshift FR II radio galaxies have, in most cases, strong emission lines and are classified as AGNs \citep{Wierzbowska2011} with peculiar optical morphologies, e.g. tails, bridges, and shells \citep{Smith1989} and bluer colors with respect to giant ellipticals. It has also been established that powerful FR IIs show a strong correlation between their radio luminosity and their optical emission-line luminosity \citep{Baum1995}, suggesting that both the optical and the radio emission originate in the same physical process.\newline
The possibility of a more general correlation between the emission lines and radio luminosities of radio AGN has been already explored in the past \citep{McCarthy1993,Zirbel1995}. Radio-loud AGNs are most likely to display emission lines in galaxies with low velocity dispersions and at radio luminosities greater than $10^{25}~WHz^{-1}$ \citep{Kauffmann2008}. A similar correlation has been observed between the emission-line luminosity and the ionization state of the gas for a sample of low-$z$ radio galaxies \citep{Saunders1989}, where the higher values of the emission-line luminosities have been measured for the more powerful radio sources, as an indication of the presence of a strong ionizing AGN-like field. However, a problem arises since the most powerful radio galaxies are generally detected at higher redshifts than the less powerful radio galaxies, due to selection effects. This makes it difficult to establish whether the correlation is between the emission-line luminosity and radio luminosity, rather than between emission-line luminosity and redshift \citep{McCarthy1993}. The well-known effect of having an increasing number of detected powerful radio AGNs with increasing redshift is not only due to a selection bias, but is is also supported by the downsizing scenario of galaxy evolution \citep[e.g.][]{Thomas2005,Cimatti2006}. Galaxies placed at higher redshift are more massive and host massive black holes that accrete producing powerful jets. These are easily detected in the radio domain, while low-redshift radio galaxies host low-mass black holes, in which only weak jets originate. The quasars detected in the radio but not in the optical regime might be heavily obscured objects. They indeed experience dust obscuration when the line of sight passes through the optically thick torus, which surrounds the black hole (unified model, Antonucci 1993).\newline
One of the current scenarios of galaxy evolution includes the possibility of a smooth transition from late-type galaxies (LTGs), which show blue colors and ongoing star formation, to red and passively evolving early-type galaxies (ETGs). This transition could be driven by the so-called AGN-feedback \citep{Cattaneo2007,Cattaneo2009}, which is thought to be responsible for star formation quenching. After the quenching, a phase of passive evolution starts, and the galaxies color turns into red. Some studies show that there is indeed a correlation between the age of galaxy stellar populations and the AGN activity \citep[][Vitale et al. 2012, in prep.]{Tortora2009}, %,)
with the oldest stars inhabiting the AGN-like galaxies. Moreover, AGNs are found to reside almost exclusively in massive galaxies with structural properties similar to normal early-type systems \citep{Heckman2004}, which are dominated by old stellar populations.\newline
The question of whether the AGN-feedback is also capable of triggering star formation in the vicinity of the black hole still has to be answered. The standard accretion mode of AGNs, which is associated with quasar activity \citep{Shakura1973}, is related to star formation in the host galaxies \citep{Kauffmann2003a}. This scenario seems to contrast with the quenching of the starburst activity due to AGN-feedback. Star formation and accretion around the black hole, both fueled by the same material, might occur with delays with respect to each other \citep{Wild2010,Tadhunter2011} and come to an end once the gas is exhausted. A fraction of those AGNs that accrete in `quasar´ mode show powerful radio jets and are therefore radio-loud.\newline
As an observational proof of this scenario, \citet{Ivezic2002} have shown that a sample of optically unresolved radio sources from the FIRST survey has bluer colors than do other SDSS objects, and Richards et al. (2002) find that the SDSS quasar candidates - which are likely to have a radio counterpart - display blue colors, especially at $z>1$, indicative of ongoing star formation. To understand the importance of the AGN-feedback in the current scenarios of galaxy evolution, a study of the AGN populations, the properties of the host galaxies, and the physical mechanisms that trigger the production of the emission lines or the radio-activity, is necessary.\newline
AGNs can be selected from a spectroscopic survey using some optical emission line ratios. Emission-line diagnostic diagrams have been extensively used during past decades to point out the connection between the galaxy nuclear activity, its morphological type \citep{Ho1997}, and its evolutionary stage \citep{Hopkins2006}. The Baldwin-Phillips-Terlevich (BPT) diagnostic diagram \citep{Baldwin1981} and its subsequent versions \citep{Veilleux1987,Kewley2003,Lamareille2004,Dopita2002,Groves2004,Groves2004b} make use of emission line ratios whose strength is a function of the hardness of the ionizing field of the galaxy, the ionization parameter $U$, and the metallicity (see Sect. 3.4). Higher ratios are thought to mainly be the product of the ionization that arises due to accretion around the black hole (which implies the AGN at the center of the galaxy is in its active phase) rather than photoionization by hot massive OB stars. This diagnostic technique, largely used in the optical wavelength regime, allows differentiation of galaxies that show activity in their nuclei and starbursts.\newline
Some attempts to link optical and radio properties of a large sample of galaxies by using combined spectroscopic and photometric information have been already made in the past. Obric et al. (2006) find strong correlations between the fraction of detected AGNs at other wavelengths and optical properties such as flux, colors, and emission-line strengths. \citet{Ivezic2002} discuss the optical and radio properties of $\sim30\ 000$ FIRST \citep{Becker1995} sources positionally associated with a SDSS source by analyzing their colors, while \citet{Best2005} compare the optical survey to both FIRST and NVSS radio surveys in order to derive the local radio luminosity functions of radio-loud AGNs and star-forming galaxies. In the SDSS Early Data Release \citep{Ivezic2002}, $\sim70$\% of FIRST sources do not have an optical counterpart within $3\arcsec$. This is probably because the majority of unmatched FIRST sources, detected down to $1$ mJy sensitivity, are too optically faint to be detected in the SDSS images. Moreover, the fraction of quasars in the FIRST catalog seem to be a strong function of the radio flux, monotonically decreasing from bright radio sources towards the FIRST radio sensitivity limit.\newline
SDSS spectra have been used to compute the line strength of several emission lines to classify galaxies into starbursts and AGNs by using diagnostic diagrams. In \citet{Ivezic2002}, the number of radio galaxies classified as AGNs rather than starbursts ($\sim30\%$) is six times more than the corresponding number for all the SDSS galaxies. Furthermore, the radio emission from AGNs turns out to be more concentrated than the radio emission from starburst galaxies, as we expect from the nuclear origin of AGN emission. Radio emission is point-like in compact quasars detected at high redshift, while local galaxies tend to have larger radio sizes. This suggests that a significant amount of the radio emission either originates outside the nuclear region or that the radio lobes are resolved. Depending on the beam size, it can happen that not all the light from the galaxy is taken into account. The relative number of AGNs decreases with the radio flux, and this is consistent with the differences in the radio luminosity functions of starburst and AGNs \citep{Machalski2000,Sadler2002}.\newline
In a multiwavelength approach, we made use of both optical - for the part concerning emission lines and diagnostic diagrams - and radio data to conduct a statistical study of the prospects of identifying radio galaxies in some well-defined regions of the low-ionization emission lines diagnostic diagrams by using a combination of radio and optical properties. A comparison of the spectroscopic measurements with some photoionization and shock models is then presented to shed light on the origin of the emission lines in AGNs and star-forming galaxies.\newline 
The paper is organized as follows: In Sect. 2 we present the optical and radio samples and their cross-matching. We describe the way we obtain the final sample and its completeness. In Sect. 3 we present our results. We exploited the diagnostic diagrams for the cross-matched sample, looking for a trend with $L_{20~cm}/L_{H\alpha}$ and redshift. We compare the spectroscopic measurements with photoionization and shock models. Discussion on the results is provided in Sect. 4. Main findings and conclusions are in Sect. 5.

\section{Optical and radio samples}

\subsection{Sloan Digital Sky Survey}
The Sloan Digital Sky Survey (SDSS) is a photometric and spectroscopic survey that covers one-quarter of the celestial sphere in the north Galactic cap \citep{York2000,Stoughton2002}. It uses a dedicated $2.5$ m wide-angle optical telescope at Apache Point Observatory in New Mexico, the United States. The spectra have an instrumental velocity resolution of $\sigma \sim 65$ km/s in the wavelength $3800-9200~$\AA $~$range. The identified galaxies have a median redshift of $z\sim0.1$. Spectra are taken with $3\arcsec$ diameter fibers ($5.7$ kpc at $z\sim0.1$), which makes the sample sensitive to aperture effects (low-redshift objects are likely to be dominated by nuclear emission, e.g. Kewley et al. 2003). Several physical parameters of the galaxies in the SDSS have been derived and listed in publicly available catalogs.\newline
The Max-Planck-Institute for Astrophysics (MPA)-Johns Hopkins University (JHU) Data Release 7 (DR7) of spectrum measurements (http://www.mpa-garching.mpg.de/SDSS/DR7/) contains the derived galaxy properties from the MPA-JHU emission line analysis for the SDSS DR7 \citep{Abazajian2009}. It represents a significant extension in size ($\sim2$ times, from $567\ 486$ to $922\ 313$ objects) and a general improvement over the previous DR4 data set. This is due to improvements both in the reduction pipeline of the SDSS and in the analysis pipeline. In addition, new stellar population synthesis spectra for the stellar continuum subtraction \citep[updated][]{Bruzual2003} are used. One caveat of the MPA-JHU DR7 is that it does not differentiate between narrow and broad emission lines. Since the narrow component alone is used in the diagnostic diagrams, this introduces a bias. An underestimation of the flux could lead to galaxies being misclassified. \citet{Ossy2011} estimate that $1.3$\% of the galaxies in the SDSS DR7 present broad components in their spectra. They also show that when a double Gaussian fit (for narrow and broad components of the same line) is performed it is possible to recover the misclassified objects and increase the number of sources in the AGN region of the diagnostic diagrams.

\subsection{Faint Images of the Radio Sky at Twenty-centimeters survey}
The Faint Images of the Radio Sky at Twenty-Centimeters Survey \citep[FIRST;][]{Becker1995} makes use of the Very Large Array (VLA) in the B-array configuration to produce a map of the $20$ cm ($1.4$ GHz) sky emission with a beam size of $5\farcs4$ and an rms sensitivity of about $0.15$ mJy/beam. The survey covers an area of about $10\ 000~\rm deg^2$ in the north Galactic cap, corresponding to the sky regions investigated by SDSS, and observed $\sim10^6$ sources. At the $1$ mJy source detection threshold, about one third of the FIRST sources show resolved structures on scales of $2\arcsec-30\arcsec$ \citep{Ivezic2002}. The FIRST catalog gives information about the continuum flux density peak ($F_{\rm peak}$) and the integrated flux density ($F_{\rm int}$) at $20$ cm, which allow separating resolved from unresolved sources. 

\subsection{Sample selection and crosscorrelation}
The SDSS DR7 spectroscopic sample makes it possible to conduct a large statistical study on a large sample of galaxies (up to $\sim10^6$ objects). With the photometric and spectroscopic information publicly available, the use of such diagnostic tools, such as the emission-line diagnostic diagrams, opens the door to new scientific interpretations. A crosscorrelation with the FIRST survey provides us with a large optical-radio sample of galaxies.\newline
Galaxies at very low redshift have angular sizes larger than the size of the fiber used for observations (for the SDSS, $3\arcsec$), so part of the emission coming from the outer regions of the galaxies could be missed. To take the aperture effect into account and to follow \citet{Kewley2003}, we consider only the objects with $z>0.04$.\newline 
Seyfert galaxies usually have bright emission lines, and they are all expected to be classified in the diagnostic diagrams. On the other hand, LINERs present much weaker emission lines. Therefore, by applying any cut in EW when selecting galaxies for our sample, we are likely to lose a part of these optically-weak emitters. In order not to be biased against LINERs and to represent as many of them as possible in the diagnostic diagrams, we opted not to apply any EW cut. Instead, we included in our sample only those galaxies whose error on the measurement of the EW of the lines that appear in the diagnostic diagrams is less than $30$\%. \newline
Our choice was supported by several tests on the cross-matched sample. The weakest emission line has been found to be the {[O\sc i]}$\lambda 6300$ line. If we keep our error cut and additionally apply a more severe cut in the measured EW of this line (EW$>3$\AA), we are left with very few galaxies ($1\ 808$). The second most affected line is {[O\sc iii]}$\lambda 5007$ ($6\ 138$ galaxies left), followed by H$\beta$ ($7\ 573$ galaxies left). The possibility of setting this condition for all three weakest emission lines has not been considered, since it would imply working with a rather small sample of radio emitters ($1\ 725$ galaxies) and poor statistics. The same cut on the strongest lines - H$\alpha$, {[N\sc ii]}$\lambda 6583$ and {[S\sc ii]}$\lambda\lambda 6717,6731$ - does not exclude a significant number of galaxies. Moreover, in a cross-matched sample where we only apply the redshift cut ($34\ 733$ galaxies), the mean logarithmic values of the emission line ratios are higher (log {[N\sc ii]}/H$\alpha=-0.13$, log {[S\sc ii]}/H$\alpha=-0.29$, {log [O\sc i]}/H$\alpha=-0.91$) than when we also apply the $30$\% error cut on all the emission lines ($9\ 594$ galaxies, {log [N\sc ii]}/H$\alpha=-0.24$, log {[S\sc ii]}/H$\alpha=-0.42$, log {[O\sc i]}/H$\alpha=-1.22$), indicating in the second case a shift in the classification from LINERs to starburst galaxies. Both H$\beta$ and {[O\sc iii]}$\lambda 5007$ appear in all the diagnostic diagrams. Their ratio is strongly affected by the weakness of the components, especially the H$\beta$ line. This line is often superimposed on an absorption component, making its strength dependent on the quality of the stellar continuum subtraction. In the case of a sample in which a cut in redshift and EW error has been applied to all the emission lines, as well as the requisite EW$>3$ \AA\ for the H$\beta$ and {[O\sc iii]}$\lambda 5007$ lines ($5\ 298$ galaxies), the line ratios slightly decrease further, moving the bulk of the population to the SF region. This is probably due to the lack of LINERs. The log {[O\sc iii]}/H$\beta$ ratios increases from $-0.01$ of the precedent case to $0.12$ instead, becoming more characteristic of Seyfert galaxies. This leads to the conclusion that we miss some of the weakest optical emitters (e.g. LINERs) when we apply an EW error cut on all the emission lines, but we are still left with a significant number of sources. In \citet{Kewley2006}, the authors prefer to consider an S/N$>3$ to be certain that the quality of the galaxy spectra is high enough to make accurate line measurements. \citet{Stasinska2006} mention that this cut does not change the visual shape of the galaxy distribution in the diagnostic diagrams, but it reduces the proportion of the objects in the right wing of the seagull.\newline
After applying redshift and EW error cut ($z>0.04$, error on the EW measurements of all the lines that are used in the diagrams is less than $30$\%), the resulting sample of SDSS galaxies with emission line measurements from DR7 contains $79\ 919$ galaxies, which is about how many objects have been used by \citet{Kewley2006}. In that case, the authors made use of a previous SDSS data release (DR4) containing fewer sources, thus our selection criteria are more restrictive. Table \ref{ew} contains the statistics on the EW of the lines for the objects that have been included in our final cross-matched sample. \newline
 \begin{table}
  \caption{\label{ew}Statistics on the EW of the lines.}
  \centering
    \begin{tabular}{|c|c|c|c|c|}
    \hline
     Emission line&EW$_{m}$&ERR$_{abs,~m}$&ERR$_{rel,~m}$\\
    \hline
     H$\alpha$&40.56&0.43&0.01\\
     H$\beta$&8.12&0.32&0.07\\
     {[O\sc iii]}$\lambda 5007$&15.53&0.36&0.08\\
     {[O\sc i]}$\lambda 6300$&2.21&0.22&0.14\\
     {[N\sc ii]}$\lambda 6583$&19.42&0.31&0.02\\
     {[S\sc ii]}$\lambda\lambda 6717,6731$&13.30&0.52&0.06\\
    \hline
    \end{tabular}
   \tablefoot{Col. 2: Mean values of the EW of the lines in \AA. Col. 3 and 4: Absolute and relative mean errors on the EW measurements.}
 \end{table}
For generating the cross-matched FIRST/SDSS sample, we used the matching results provided by the SDSS DR7 via Casjobs \citep{OMullane2005}, based on a matching radius of $1\arcsec$. The resulting sample contains $37\ 488$ radio emitters and represents nearly $40\%$ of the $\sim10^5$ FIRST sources and around $4\%$ of the $\sim10^6$ SDSS sources. SDSS spectra are available for all objects of the matched sample. Some objects have been lost during the matching procedure owing to the centering problems of radio emission with respect to the optical emission, especially in the case of extended sources. Moreover, \citet{Best2005} find that radio galaxies extracted from the main spectroscopic sample of the SDSS reside in very massive early-type galaxies, with weak or undetectable optical emission lines. This further proves that the cross-matching process implies the loss of many radio emitters that do not have optical counterparts with emission lines.\newline
Another considerable number of radio emitters that have an optical counterpart are lost after applying the redshift cut ($z>0.04$) and the error cut on the EW of the lines involved in the diagnostic diagrams. Our final cross-matched sample consists of $9\ 594$ objects, which corresponds to $25.6$\% of the full cross-matched optical-radio sample, $\sim 9.6$\% of the original radio sample (FIRST) and $\sim1$\% of all the galaxies in the MPA-JHU data release.\newline
The radio luminosity has been calculated from the integrated flux provided by the FIRST catalog, assuming a cosmology with $H_0=70$ km s$^{-1}$ Mpc$^{-1}$, $\Omega_M=0.3$, $\Omega_{\Lambda}=0.7$.

\subsection{Completeness of the sample}
\begin{figure*}
 \centering
  \includegraphics[width=17.5cm]{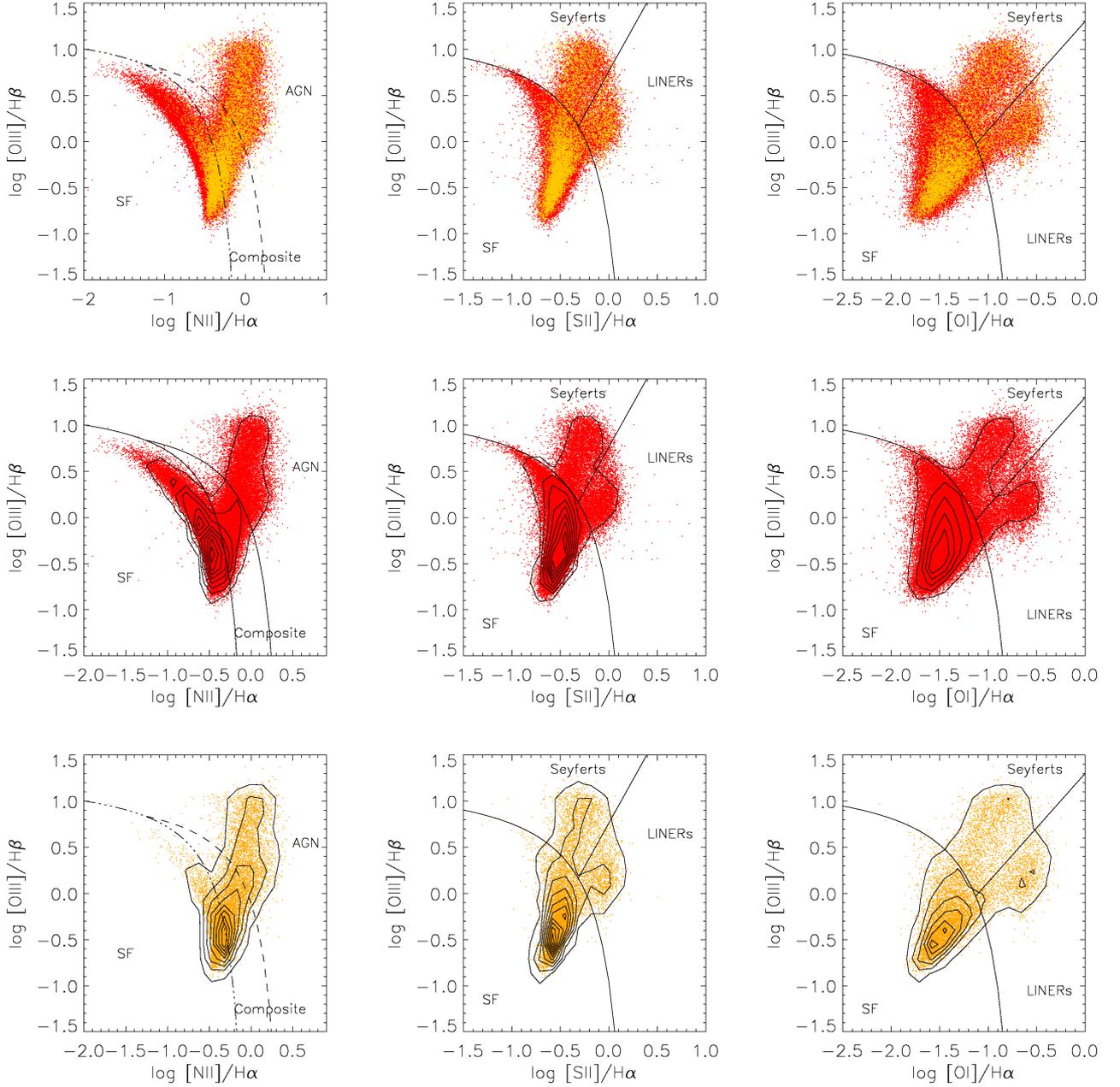}
%I applied the Kewley cut to both samples
 \caption {\label{bpt} Optical emission-line diagnostic diagrams. The SDSS (DR7) targets are represented in red, while the radio emitters from the cross-matched sample are plotted in orange. Demarcation curves in the left panels ({[N\sc ii]} diagram) are by \citet{Kewley2001} (dashed) and \citet{Kauffmann2003a} (dot-dashed); in the middle ({[S\sc ii]} diagram) and right ({[O\sc i]} diagram) panels, the demarcation curves are by \citet{Kewley2006}. The top panels show the SDSS targets overplotted with the radio emitters. The middle and bottom panels show the density function of our data distribution. Density levels represent $700$ galaxies  per contour in case of SDSS targets, while it is $70$ for the radio emitters.}
\end{figure*}
 \begin{table*}
  \caption{\label{statistics}Statistics on the SDSS DR7 sample and our optical-radio sample including the number of different objects placed in the AGN, composite, or SF region (left plots, Fig.\ref{bpt}), or Seyfert, LINER, or SF region (middle and right plots) of the diagnostic diagrams.}
  \centering
    \begin{tabular}{|c|c|c|c|}
    \hline
     Diagram&Spectral type&SDSS DR7&Cross-match\\
    \hline
     &AGNs&11 756$\pm$108 (14.8\%)&2 968$\pm$54 (31.5\%)\\
     {[N\sc ii]}&Composites&13 856$\pm$118 (17.5\%)&3 021$\pm$55 (32.1\%)\\
     &SFGs&53 623$\pm$232 (67.7\%)&3 419$\pm$58 (36.3\%)\\
    \hline
     &Seyferts&7 646$\pm$87 (9.6\%)&1 747$\pm$42 (18.6\%)\\
     {[S\sc ii]}&LINERs&4 107$\pm$64 (5.2\%)&1 012$\pm$32 (10.8\%)\\
     &SFGs&67 482$\pm$260 (85.2\%)&6 649$\pm$81 (70.7\%)\\    
    \hline
     &Seyferts&10 269$\pm$101 (13.0\%)&2 180$\pm$47 (23.2\%)\\
     {[O\sc i]}&LINERs&6 543$\pm$81 (8.3\%)&1 500$\pm$39 (15.9\%)\\
     &SFGs&62 423$\pm$250 (78.8\%)&5 728$\pm$76 (60.9\%)\\
    \hline
    \end{tabular}
   \tablefoot{The relative fraction of different spectral types is reported in parenthesis.}
 \end{table*}
Our sample completeness is strongly related to the completeness of the parent samples (SDSS and FIRST). The SDSS spectroscopic sample can be considered as complete, in the sense that the biggest incompleteness comes from galaxy misclassifications owing to mechanical spectrograph constraints \citep[$6$\%,][]{Blanton2003}, which causes slight underrepresentation of high-density regions. For a few sources ($<1$\%), the redshift cannot be determined or it has been derived in an incorrect way. Moreover, a few targets ($\sim1$\%) are contaminated by galactic stars. According to \citet{Blanton2006}, the mechanical constraints are related to the impossibility of placing the fibers close enough to each other. When two galaxies are found to have a small separation, only one source is chosen (independently from its magnitude or surface brightness), and this does not result in a luminosity bias. Reliability and accuracy of the catalog of radio sources extracted from the FIRST images is discussed in \citet{White1997}.

\section{Results}
A statistical study of the spectral properties of radio emitters from the SDSS has been carried out starting from the determination of the fraction of star-forming galaxies and AGNs composing our sample. For this purpose, we exploited the emission-line diagnostic diagrams to separate sources with a hard ionized spectrum (AGNs) and those that are dominated by star formation, characterized by a considerably softer ionizing field. Since all the emission lines that are used in the diagrams are placed close to each other in the spectrum, their ratios are almost insensitive to reddending, so we do not need to correct for the interstellar and galactic extinction. The luminosity of the H$\alpha$ line has been derived after correcting the corresponding flux for the visual extinction by using a theoretical H$\alpha/$H$\beta$ Balmer ratio of $2.86$.
\begin{figure*}
  \centering
  \includegraphics[width=11cm]{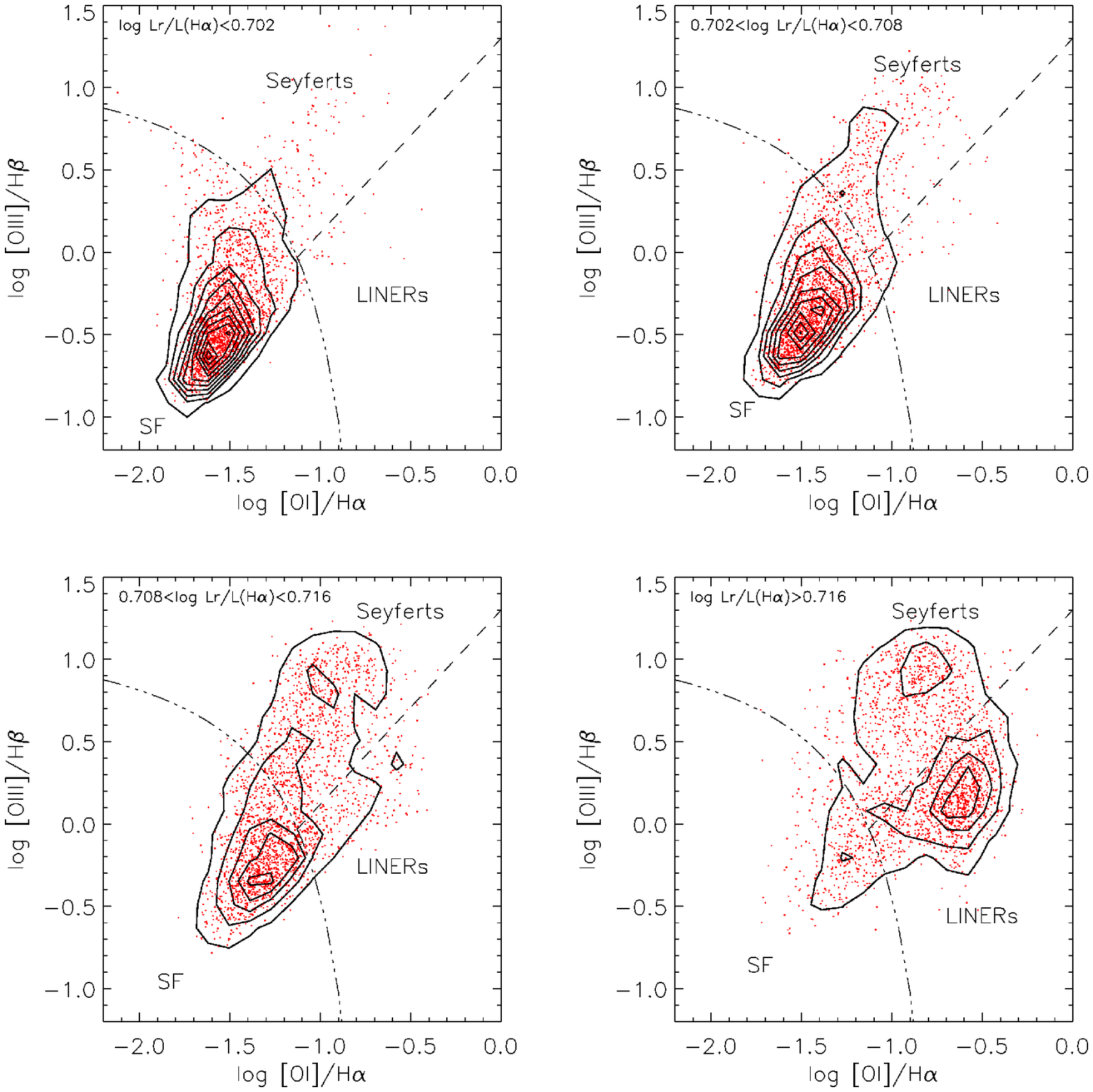}
  \caption{{[O\sc i]}-based diagnostic diagrams for the optical-radio sample. From top to bottom and from left to right, log$(L_{20~cm}/L_{H\alpha})$ increases. The number of radio emitters per bin is constant and equal to $2\ 350\pm25$. The contours represent the number density of the radio emitters ($20$ galaxies per density contour). The bulk of the population shifts from the SF region to the composite or AGN region with increasing $L_{20~cm}/L_{H\alpha}$. The transition seems to occur for log$(L_{20~cm}/L_{H\alpha})>0.716$.}
 \label{oi_lum_tresh}
 \end{figure*}
 \begin{figure*} 
 \centering
  \includegraphics[width=11cm]{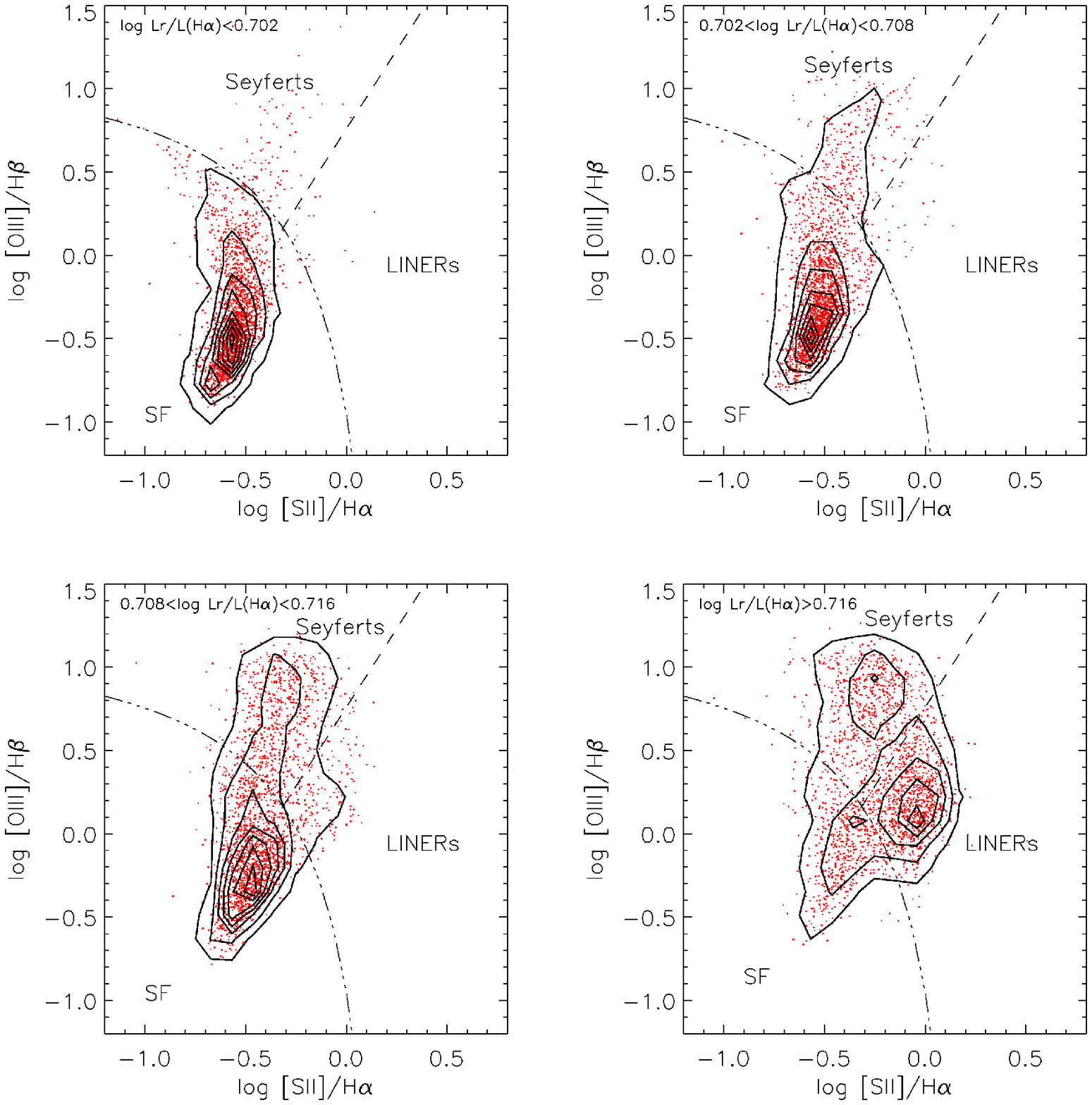}
  \caption{As in Fig.\ref{oi_lum_tresh} but for for the {[S\sc ii]}-based diagnostic diagram. The contours represent the number density of the radio emitters ($40$ galaxies per density contour in the upper panels, $20$ in the lower panels).}
 \label{sii_lum_tresh}
 \end{figure*}
\begin{figure*} 
 \centering
  \includegraphics[width=11cm]{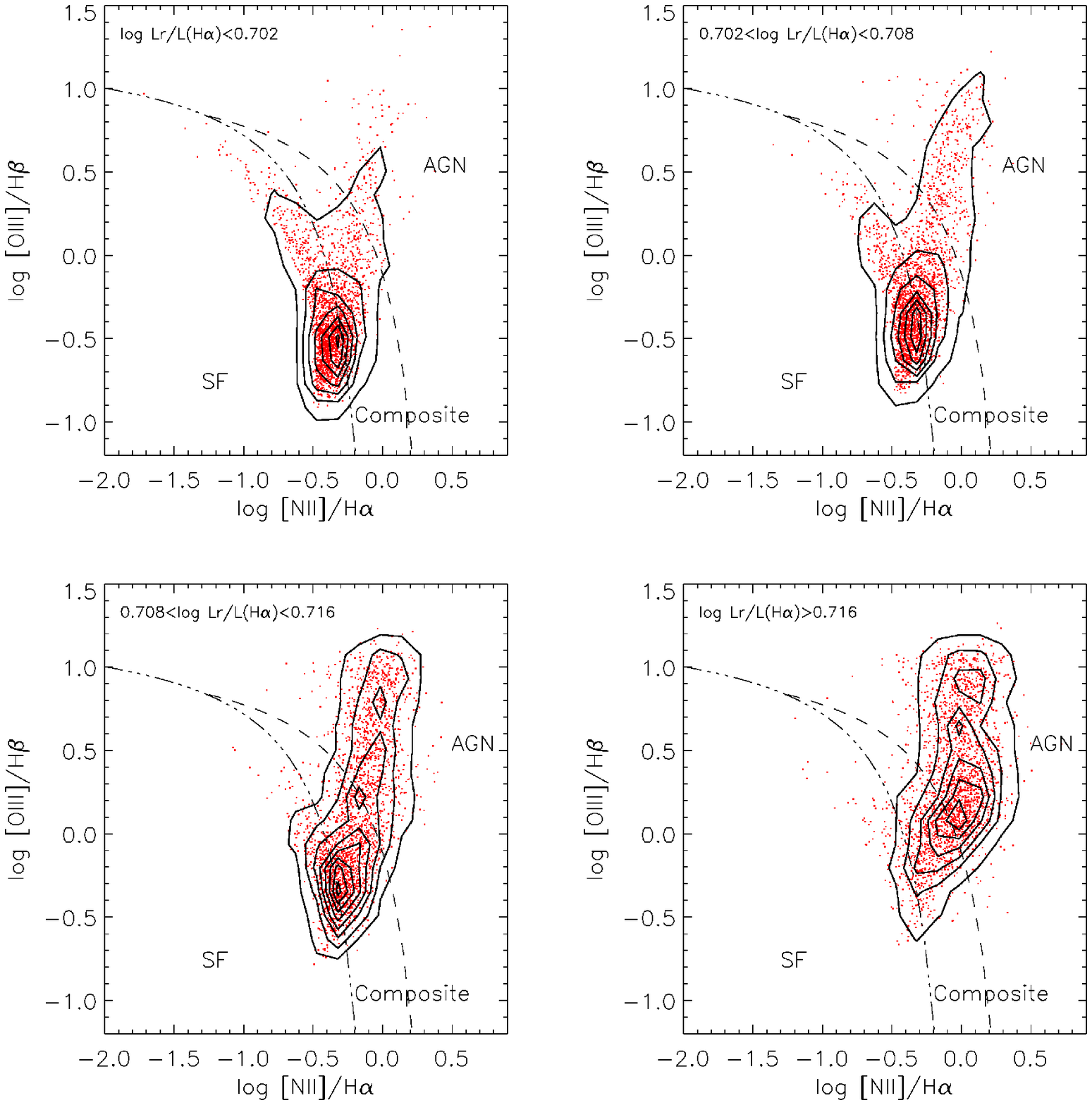}
  \caption{As in Fig.\ref{oi_lum_tresh} but for for the {[N\sc ii]}-based diagnostic diagram. The contours represent the number density of the radio emitters ($40$ galaxies per density contour in the upper panels, $20$ in the lower panels).}
 \label{nii_lum_tresh}
 \end{figure*}

\subsection{Diagnostic diagrams for the SDSS and the cross-matched sample}
Figure \ref{bpt} presents the optical emission-line diagnostic diagrams for the MPA-JHU and the cross-matched radio-optical samples. In all the diagrams, $79\ 235$ of the $79\ 919$ (99.1\%) galaxies in our MPA-JHU sub-sample are represented, and $9\ 408$ of the $9\ 594$ radio emitters (98,1\% of the optical-radio sample). The difference comes from the lack of some emission-line measurements in the parent catalog of SDSS galaxies and especially the {[O\sc i]} line, which is weaker than the other lines. \newline
The dashed demarcation curve in the plots of the first column of Fig. \ref{bpt} ({[N\sc ii]} or BPT diagram) has been derived by \citet{Kewley2001} by constructing a detailed continuous starburst model with broad realistic metallicity and ionization parameter ranges, in order to find an upper limit for the position of the star-forming galaxies in the diagram. This upper limit can be used to separate AGNs from star-forming galaxies. The dot-dashed curve has been derived by \citet{Kauffmann2003a}, who propose an empirical and more conservative cut to identify starbursts by using the large sample of emission line galaxies in the SDSS. This method selects many fewer star-forming galaxies than AGNs - according to the main ionizing mechanism that produces the emission lines - compared to the Kewley's criteria. The enclosed region between the two curves is considered to be populated by mixed or transitional objects. In the middle ({[S\sc ii]} diagram) and right ({[O\sc i]} diagram) panels, the demarcation curves are from \citet{Kewley2006}. Another line allows us to separate Seyfert galaxies from LINERs.\newline
In Table \ref{statistics} we present the statistics on the samples. For each diagnostic diagram, we report the number of AGNs (with the distinction between Seyferts and LINERs, when available), composites, and star-forming galaxies for both samples. The absolute number of objects per region is followed by the purely statistical error (Poisson error estimated as $\sqrt{N}$) and the percentage with respect to the total number of galaxies with measured emission-lines. A higher relative number of radio emitters are placed in the transitional or AGN regions than in the full optical sample, and the percentage of radio emitters in the left arm of the `seagull´ shape defined by the SDSS targets ({[N\sc ii]} diagram) is lower. The SDSS is strongly dominated by star formation rather than AGN-like emission. Figure \ref{bpt} shows that the AGNs and the radio galaxies with AGNs are drawn from a population that has higher metallicity than the overall SDSS sample. The lower abundance objects are predominantly star-forming, and populate the upper left-hand portions of these diagrams. The {[O\sc i]} diagram is the one that shows the largest number of objects classified as AGNs ($39.1$\% of the cross-matched sample).\newline
We should take into account that the redshift we are considering in our study is limited because of the choice of the diagnostic diagrams. Some of the emission lines are shifted out of the SDSS spectral range for objects with $z\gtrapprox0.4$ ($\sim2$\% of the optical-radio sample).

\subsection{Diagnostic diagrams for the cross-matched sample: Trend with $L_{20~cm}/L_{H\alpha}$}
\begin{figure*}
  \centering
  \includegraphics[width=14cm]{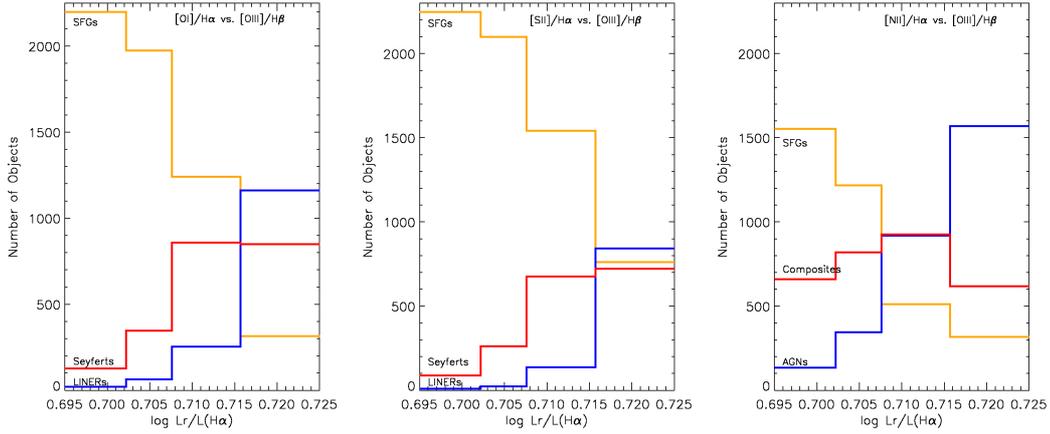}
  \caption{\label{histogram}Number of identified spectral types in the $L_{20~cm}/L_{H\alpha}$ bins per each diagnostic diagram. Star-forming galaxies are in orange, Seyferts in red, and LINERs in blue (composites are in red and AGNs in blue in the {[N\sc ii]}-based diagram). The $L_{20~cm}/L_{H\alpha}$ value is indicated on the x-axis.}
 \end{figure*}

The luminosity of the H$\alpha$ line, $L_{H\alpha}$, is considered to be a good optical star formation rate (SFR) indicator \citep{Moustakas2006}. The ratio between the radio luminosity and $L_{H\alpha}$ can be used to compare emission from radio components with the emission from young stars. We divided the cross-matched optical-radio sample into four bins with increasing $L_{20~cm}/L_{H\alpha}$ containing approximately $2\ 350$ galaxies each to keep the number of objects per bin constant and to search for a threshold above which the objects start to be classified as AGNs (Seyfert or LINERs) rather than star-forming galaxies.\newline
We would expect to find a correlation between the optical and the radio emission, according to the hypothesis that they are linked to each other (e.g. $L_{20~cm}-L_o$ relation, where $L_o$ is the luminosity in the optical) because they come from the same physical process. Strong emission lines, such as {[O\sc iii]}$\lambda5007$, {[O\sc ii]}$\lambda3727$, and H$\beta$, are indeed believed to come from powerful radio emitters \citep{Baum1989,Rawlings1989,Rawlings1991,Morganti1992, Zirbel1995,Tadhunter1998,Best2012}. These lines are also used to select AGNs from a spectroscopic survey, pointing to a possible correlation between the AGN-detection rate and the radio luminosity of the host galaxies.\newline
We found that the peak of the distribution of the radio emitters shifts from the SF region of the diagrams to the composite or AGN part (on the right-hand side of the diagrams) for increasing log$(L_{20~cm}/L_{H\alpha})$ (Figs. \ref{oi_lum_tresh}, \ref{sii_lum_tresh}, \ref{nii_lum_tresh}). In particular, the upper left-hand panels mostly show starbursts with high metallicity, while the bottom left-hand panels display a mixed population and the bottom right-hand panels show a nearly pure AGN population, together with some metal-rich starbursts. The distribution shows a peak in the LINER region for log$(L_{20~cm}/L_{H\alpha})>0.716$, where $50$\% of the radio emitters are classified as LINERs in the {[O\sc i]}-based diagram ($36.2$\% in the {[S\sc ii]}-based diagram, see Table \ref{statistics_L_ratio}). In the upper right- and bottom left-hand panels ($0.702<$log$L_{20~cm}/L_{H\alpha}<0.708$ and $0.708<$log$L_{20~cm}/L_{H\alpha}<0.716$) of each figure, the Seyfert region appears increasingly more populated (from a few per cent up to more than $30$\%), while the number of Seyfert galaxies seems to remain constant for log$(L_{20~cm}/L_{H\alpha})>0.716$. In contrast, the increase in the number of LINERs becomes exponential in the last bin, where it goes from the few per cent of the first three bins to half of the entire population of radio emitters. All diagrams exhibit this behavior, though the trend is most obvious in the {[O\sc i]} diagram, which is also the most sensitive to shocks. In Table \ref{statistics_L_ratio} and Fig. \ref{histogram}, we report the statistics for the $L_{20~cm}/L_{H\alpha}$ bins on all the diagnostic diagrams. The number of Seyfert galaxies and AGNs in the BPT diagram increases progressively by a factor $2.7$ in the first three bins, while it remains almost constant in the last bin. The number of LINERs drastically increases ($\times5$) from the third to the last bin, while the number of SFGs decreases in the last bin. For each diagnostic diagram we indicate the number of classified star-forming galaxies, Seyferts, and LINERs (or composites and AGNs in the {[N\sc ii]}-based diagram) per $L_{20~cm}/L_{H\alpha}$ bin.
 \begin{table*}
  \caption{\label{statistics_L_ratio}Statistics on the $L_{20~cm}/L_{H\alpha}$ (in logarithmic value) bins of the optical-radio sample to show the number of different objects placed in the Seyfert, LINER, or SF region (Figs. \ref{oi_lum_tresh}, \ref{sii_lum_tresh}), or AGN, composite, or SF region (Fig. \ref{nii_lum_tresh}) of the diagnostic diagrams.}
  \centering
    \begin{tabular}{|c|c|c|c|c|c|}
    \hline
     Diagram&Spectral Type&$L_{20~cm}/L_{H\alpha}<0.702$&$0.702<L_{20~cm}/L_{H\alpha}<0.708$&$0.708<L_{20~cm}/L_{H\alpha}<0.716$&$L_{20~cm}/L_{H\alpha}>0.716$\\
    \hline
     &Seyferts&126$\pm$11 (5.4\%)&347$\pm$19 (14.5\%)&859$\pm$29 (36.5\%)&848$\pm$29 (36.5\%)\\
     {[O\sc i]}&LINERs&20$\pm$5 (0.8\%)&64$\pm$8 (2.7\%)&254$\pm$16 (10.8\%)&1 162$\pm$34 (50.0\%)\\
     &SFGs&2 199$\pm$47 (93.8\%)&1 973$\pm$44 (82.8\%)&1 241$\pm$35 (52.7\%)&315$\pm$18 (13.5\%)\\
    \hline
     &Seyferts&88$\pm$9 (3.7\%)&261$\pm$16 (10.9\%)&676$\pm$26 (28.7\%)&722$\pm$27 (31.0\%)\\
     {[S\sc ii]}&LINERs&10$\pm$3 (0.4\%)&24$\pm$5 (1.0\%)&136$\pm$12 (5.8\%)&842$\pm$29 (36.2\%)\\
     &SFGs&2 247$\pm$47 (95.8\%)&2 099$\pm$46 (88.0\%)&1 542$\pm$39 (65.5\%)&761$\pm$28 (32.7\%)\\    
    \hline
     &AGNs&134$\pm$12 (5.7\%)&346$\pm$19 (14.5\%)&918$\pm$30 (39.0\%)&1 570$\pm$40 (67.5\%)\\
     {[N\sc ii]}&Composites&659$\pm$26 (28.1\%)&820$\pm$29 (34.4\%)&925$\pm$30 (39.3\%)&617$\pm$25 (26.5\%)\\
     &SFGs&1 552$\pm$39 (66.2\%)&1 218$\pm$35 (51.1\%)&511$\pm$23 (21.7\%)&138$\pm$12 (5.9\%)\\
    \hline
    \end{tabular}
  \tablefoot{The relative fraction of spectral types is reported in parenthesis.}
 \end{table*}
 \begin{figure*}
  \centering
  \includegraphics[width=11cm]{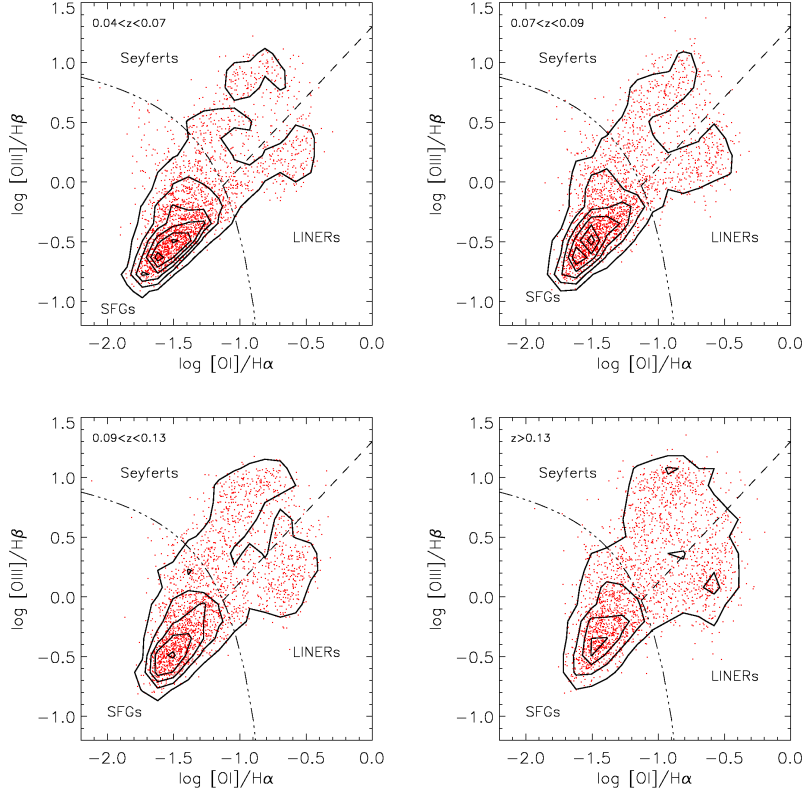}
  \caption{\label{oi_z_tresh}{[O\sc i]}-based diagnostic diagrams for the optical-radio sample. From top to bottom and from left to right, $z$ increases. The number of radio emitters per bin is constant and equal to $2\ 350\pm5$. The contours represent the number density of the radio emitters ($20$ galaxies per level). The bulk of the population is always placed in the SF region, though the number of AGNs slightly increases with $z$.}
 \end{figure*}
 \begin{figure*}
  \centering
  \includegraphics[width=11cm]{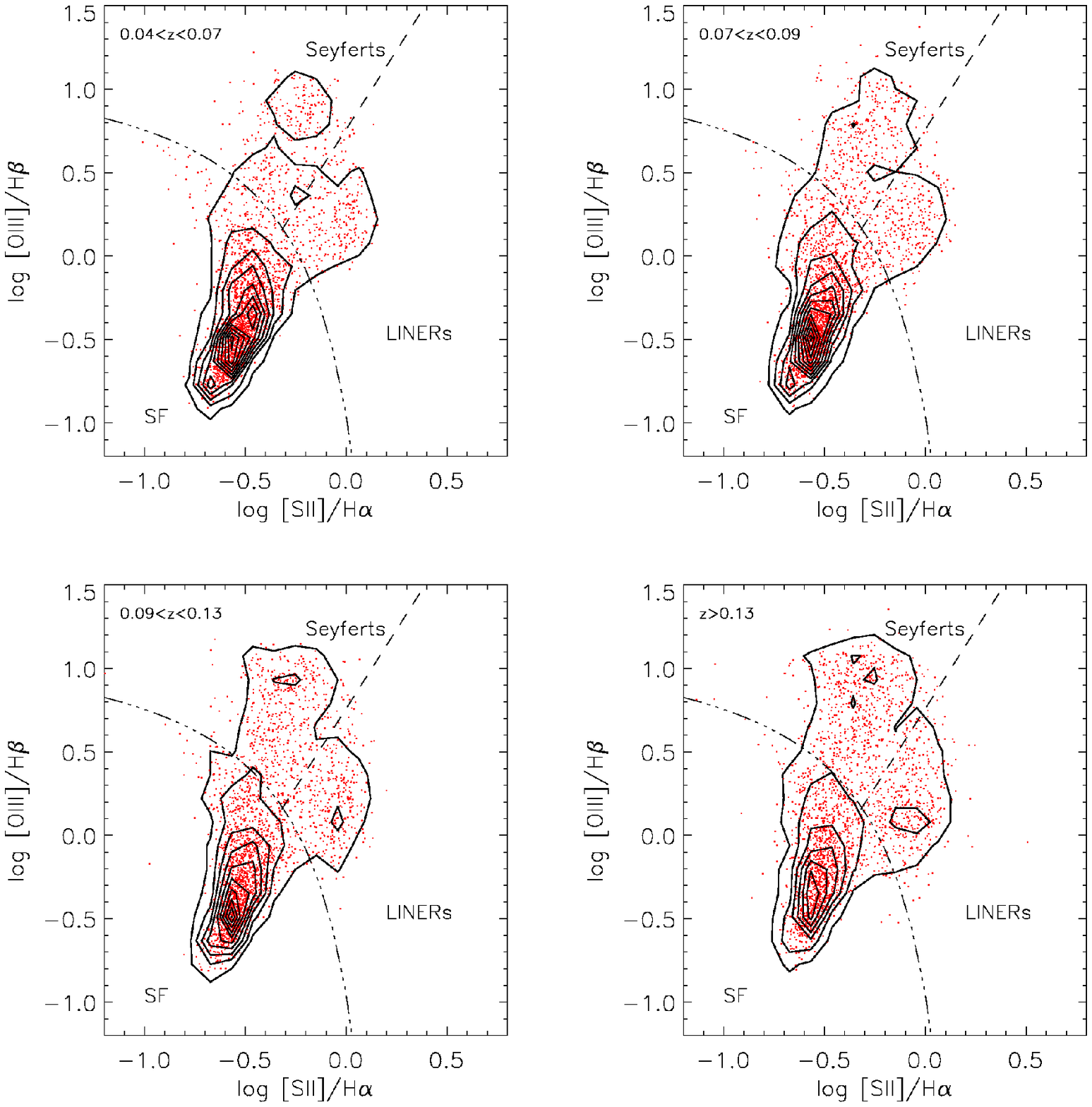}
  \caption{\label{sii_z_tresh}As in Fig.\ref{oi_z_tresh} but for the {[S\sc ii]}-based diagnostic diagram. The contours represent the number density of the radio emitters ($30$ galaxies per level in the upper panels, $20$ in the lower panels.)}
 \end{figure*}
\begin{figure*}
  \centering
  \includegraphics[width=11cm]{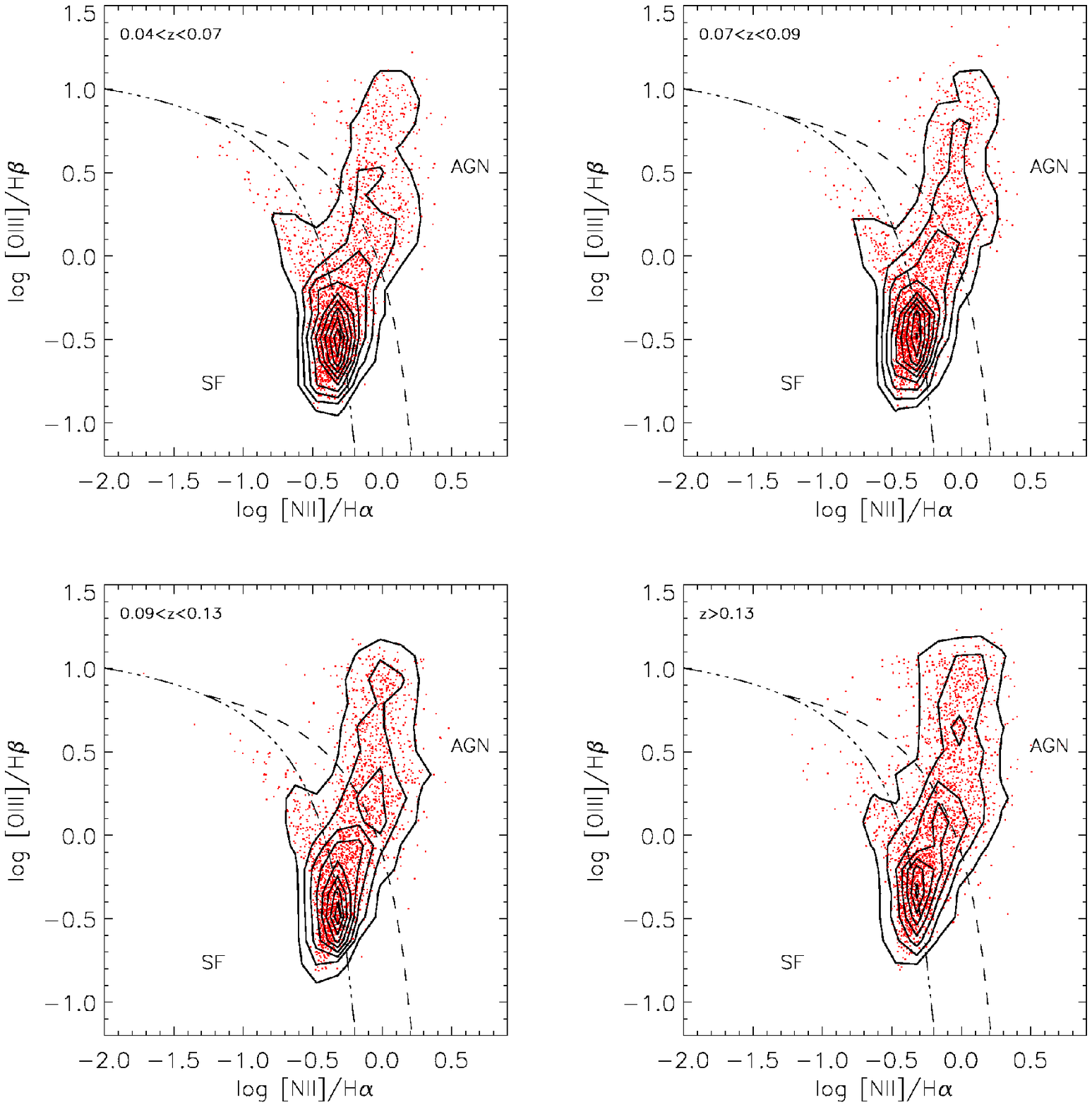}
  \caption{\label{nii_z_tresh}As in Fig.\ref{oi_z_tresh} but for the {[N\sc ii]}-based diagnostic diagram. The contours represent the number density of the radio emitters ($30$ galaxies per level in the upper panels, $20$ in the lower panels.}
 \end{figure*}

\subsection{Diagnostic diagrams for the cross-matched sample: Trend with redshift}
We exploited the diagnostic diagrams further as a function of redshift by dividing the sample into four bins containing approximately $2\ 350$ objects each. As for the $L_{20~cm}/L_{H\alpha}$ bins, this choice comes from the intent to keep the number of objects per bin constant, together with the intention to further explore the dependence of the optical classification with redshift.
\begin{table*}
 \caption{\label{statistics_z}Statistics on the z bins of the optical-radio sample showing the number of different objects placed in the Seyfert, LINER, or SF region (Figs. \ref{oi_z_tresh}, \ref{sii_z_tresh}) or AGN, composite, or SF region (Fig. \ref{nii_z_tresh}) of the diagnostic diagrams.}
  \centering
    \begin{tabular}{|c|c|c|c|c|c|}
    \hline
     Diagram&Spectral type&0.04$<z<$0.07&0.07$<z<$0.09&0.09$<z<$0.13&$z>$0.13\\
    \hline
     &Seyferts&385$\pm$20 (16.4\%)&461$\pm$21 (19.6\%)&585$\pm$24 (24.9\%)&749$\pm$27 (31.7\%)\\
     {[O\sc i]}&LINERs&288$\pm$17 (12.3\%)&314$\pm$18 (13.3\%)&386$\pm$20 (16.5\%)&512$\pm$23 (21.7\%)\\
     &SFGs&1 675$\pm$41 (71.3\%)&1 580$\pm$40 (67.1\%)&1 374$\pm$37 (58.6\%)&1 099$\pm$33 (46.6\%)\\
    \hline
     &Seyferts&308$\pm$17 (13.1\%)&364$\pm$19 (15.5\%)&465$\pm$22 (19.8\%)&610$\pm$25 (25.8\%)\\
     {[S\sc ii]}&LINERs&213$\pm$15 (9.1\%)&211$\pm$14 (9.0\%)&261$\pm$16 (11.1\%)&327$\pm$18 (13.8\%)\\
     &SFGs&1 827$\pm$43 (77.8\%)&1 780$\pm$42 (75.6\%)&1 619$\pm$40 (69.0\%)&1 423$\pm$38 (60.3\%)\\    
    \hline
     &AGNs&553$\pm$23 (23.5\%)&620$\pm$25 (26.3\%)&790$\pm$28 (33.7\%)&1 005$\pm$32 (42.6\%)\\
     {[N\sc ii]}&Composites&758$\pm$27 (32.3\%)&744$\pm$27 (31.6\%)&740$\pm$27 (31.6\%)&779$\pm$28 (33.0\%)\\
     &SFGs&1 037$\pm$32 (44.2\%)&991$\pm$31 (42.1\%)&815$\pm$28 (34.7\%)&576$\pm$24 (24.4\%)\\
    \hline
    \end{tabular}
   \tablefoot{The relative fraction of spectral types is reported in parenthesis.}
 \end{table*}
 \begin{figure*}
  \centering
  \includegraphics[width=14cm]{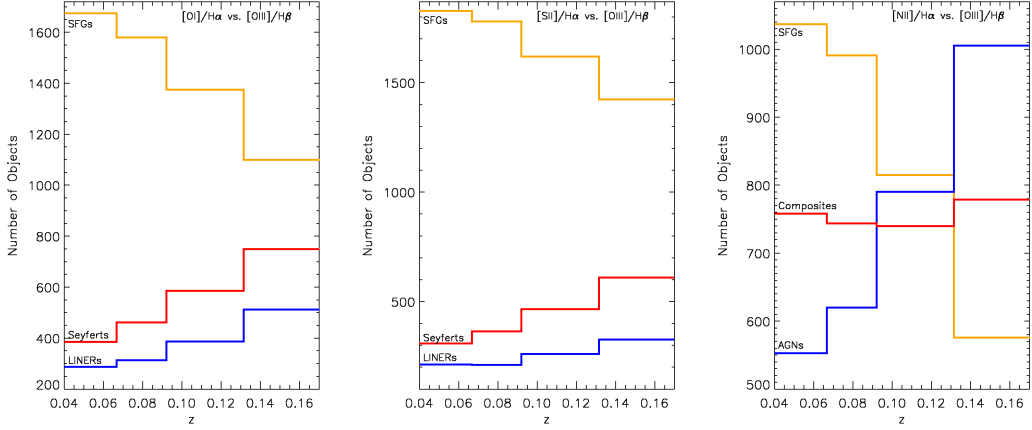}
  \caption{\label{histogram_z_trend}Number of identified spectral types in the z bins for each diagnostic diagram. Star-forming galaxies are in orange, Seyferts in red, and LINERs in blue (composites are in red and AGNs in blue in the {[N\sc ii]}-diagram). The redshift is indicated on the x-axis}
 \end{figure*}
 \begin{figure*} 
 \centering
  \includegraphics[width=16cm]{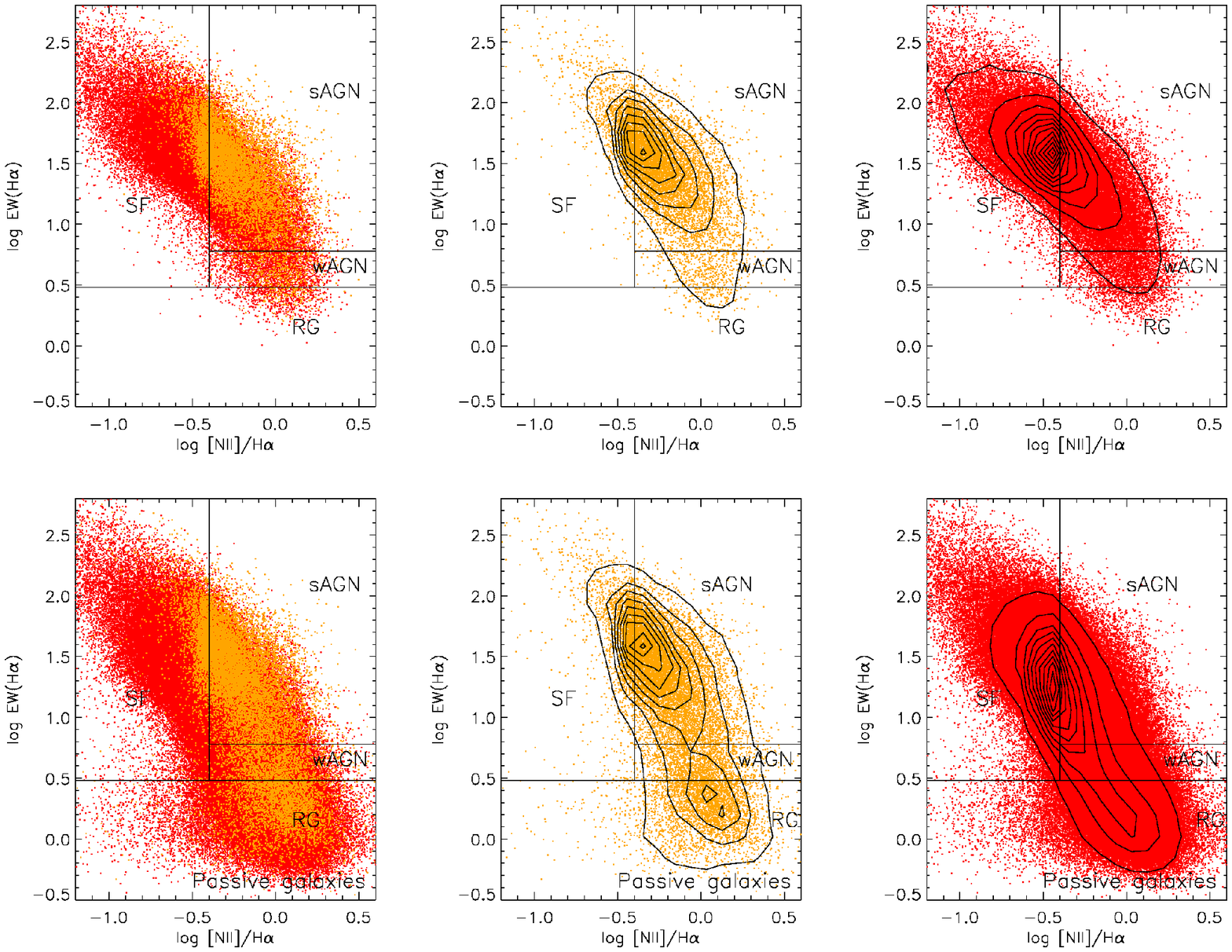}
  \caption{\label{WHAN}WHAN diagnostic diagrams for the cross-matched (orange) and the SDSS (red) samples. The upper-left panel shows the radio emitters of the SDSS superimposed on the SDSS parent sample. The lower panels represent samples where a less severe error cut (only on {[N\sc ii]} and H$\alpha$) has been applied. In this way, a considerable number of objects have been recovered, which appear to be classified as wAGNs or RGs. Density levels represent $70$ galaxies per contour in the middle panels, $500$ galaxies per contour in the upper-right panel and $2\ 000$ galaxies per contour in the bottom-right panel.}
 \end{figure*}
We found that the distribution of the radio emitters in the diagnostic diagrams depends slightly on the redshift (Figs. \ref{oi_z_tresh}, \ref{sii_z_tresh}, \ref{nii_z_tresh}). The number of AGNs (both LINERs and Seyferts) always increases with $z$. The bulk of the population remains in the SF, composite or SF+composite region in all $z$ bins (see statistics on Table \ref{statistics_z} and Fig. \ref{histogram_z_trend}). The number of Seyferts, LINERs and AGNs in the BPT diagram is nearly double in the last bin with respect to the first one. The number of composites remains almost constant, while  the number of SFGs decreases of a variable percentage ($22$\% in the {[S\sc ii]} diagram, $34$\% in the {[O\sc i]} diagram and $45$\% in the {[N\sc ii]} diagram).

\subsection{The {[N\sc ii]}/H$\alpha$ vs. equivalent width of the H$\alpha$ line (WHAN) diagram}
\begin{figure*} 
 \centering
  \includegraphics[width=12cm]{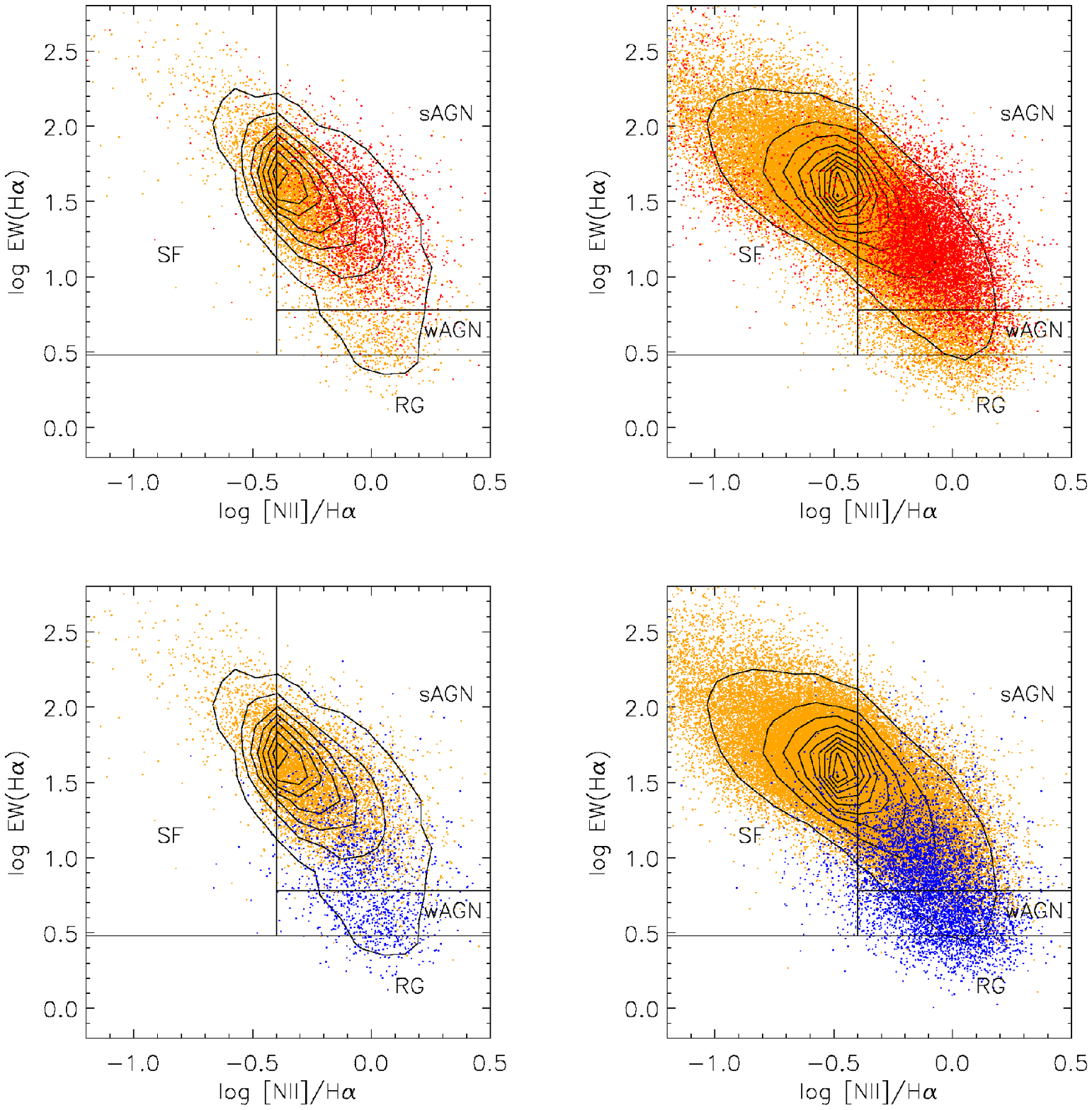}
  \caption{\label{WHAN_oi}WHAN diagnostic diagrams for the cross-matched (left column) and the SDSS samples (right column, both in orange) overplotted with AGNs selected from the {[O\sc i]}-based diagram. The upper-left plot shows the Seyferts (in red) among the radio emitters, while the upper-right plot shows the Seyferts in the parent SDSS sample. The lower panels show the LINERs distribution (in blue). Contours number density refers to the underlying orange distribution of each plot and is equal to $70$ galaxies per contour in the left panels and to $700$ galaxies per contour in the right panels.}
 \end{figure*}
\begin{figure*} 
 \centering
  \includegraphics[width=12cm]{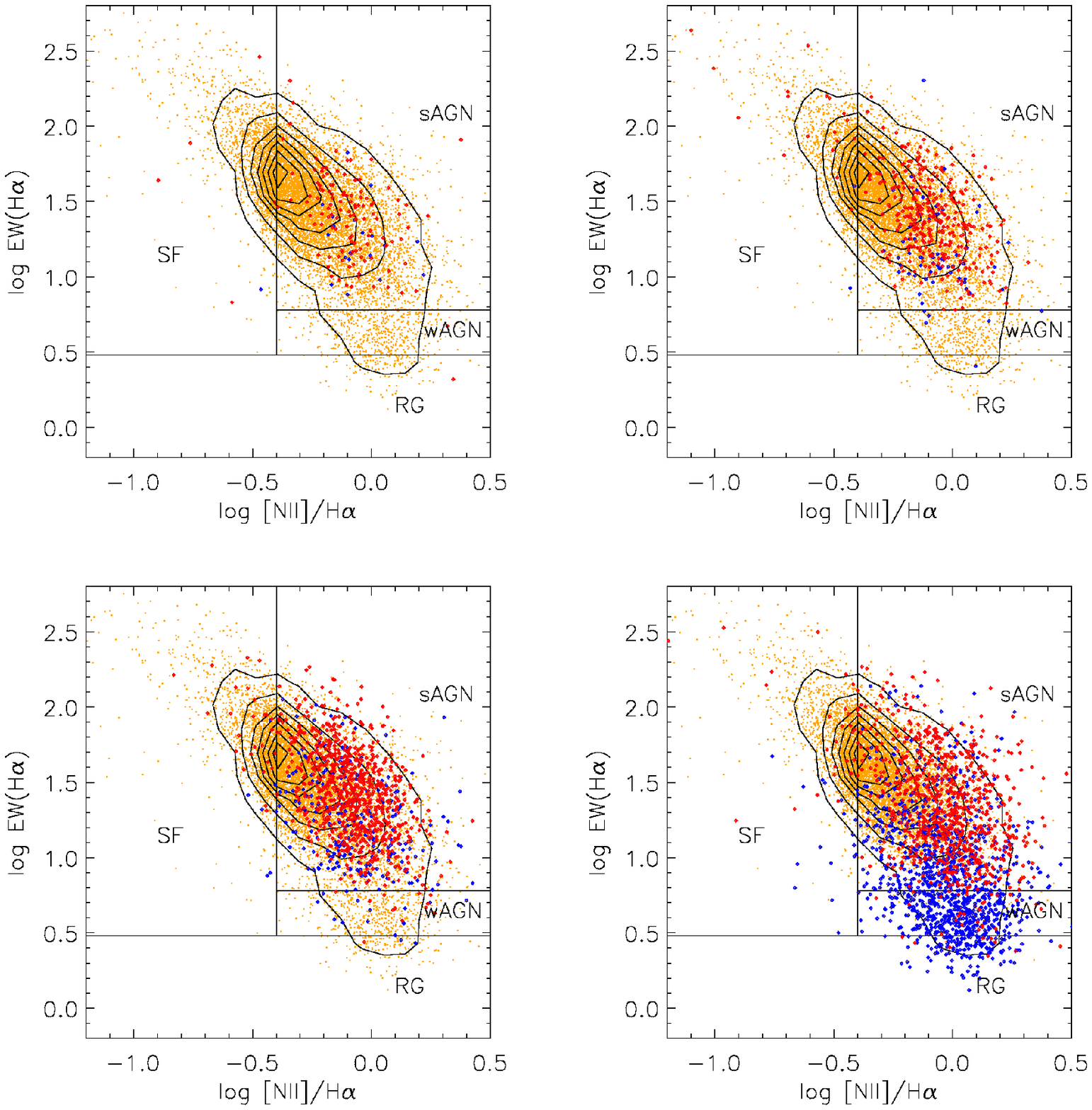}
  \caption{\label{WHAN_L}WHAN diagnostic diagrams for the cross-matched sample (orange) overplotted with Seyferts (red) and LINERs (blue) selected from the {[O\sc i]}-based diagram. Each plot represents, from the top-left to the bottom-right, a different $L_{20~cm}/L_{H\alpha}$ bin (Figs. \ref{oi_lum_tresh}, \ref{sii_lum_tresh}, \ref{nii_lum_tresh}). Contours number density refers to the underlying orange distribution of each plot and is equal to $70$ galaxies per contour.}
 \end{figure*}
Emission from some of the galaxies classified as LINERs in the diagnostic diagrams is nowadays thought to be triggered by post-asymptotic giant branch (post-AGB) and white dwarf stars, which are abundant in early-type galaxies. The radiation from old stellar populations in these galaxies, classified as `retired´ \citep[RGs,][]{Stasinska2008}, is harder than the radiation produced by young stars, providing higher emission-line ratios than those typical of star-forming regions. Because of the tight correlation between metallicity, $Z$, and ionization parameter, $U$, {[N\sc ii]}/H$\alpha$ can be used as an empirical measure of the gas metallicity up to {[N\sc ii]}/H$\alpha=0.4$, where higher values become an indication of the presence of AGN activity \citep{vanZee1998,Pettini2004,Stasinska2006}. On the other hand, the H$\alpha$ equivalent width, EW(H$\alpha$), is considered as a powerful star formation indicator, although it is also related to strong non thermal excitation (e.g. the one from Seyferts). Therefore, the {[N\sc ii]}/H$\alpha$ line ratio and EW(H$\alpha$) can be combined in a diagnostic diagram \citep[WHAN,][]{CidFernandes2010}.\newline
The WHAN diagram can distinguish between retired galaxies and true AGNs, besides having the advantage of being able to classify galaxies with weak {[O\sc iii]} and/or H$\beta$ lines. With this diagram it has been found that a large number of weak-line emitting galaxies in the SDSS have LINER-like emission with $3<$EW(H$\alpha)<6~$ \AA\ in the case of actual LINERs (labeled in the diagram as wAGN, where `w´ stands for `weak´) and EW(H$\alpha)<3~$ \AA\ for retired galaxies \citep{CidFernandes2011}. Galaxies that have both high values of EW(H$\alpha$) and {[N\sc ii]}/H$\alpha$ are classified as strong AGNs (sAGNs).\newline
In Fig. \ref{WHAN}, we show the WHAN diagram for the cross-matched optical-radio sample and for the SDSS parent sample. For the subsample used in this work, radio emitters and parent population (upper panels) exhibit a different distribution, with the former having the peak of the distribution in the sAGN region but close to the demarcation line that separates it from the star-forming galaxies. In the SF region, the number density drops and the contours show a very steep gradient. The majority of galaxies showing LINER-like emission in the classical diagnostic diagrams are found to be true LINERs rather than retired galaxies, because they are mostly placed in the wAGN zone. On the other hand, in the upper right-hand panel the distribution appears to be spread more across the SF region, with relatively fewer true AGNs (both strong and weak) and retired galaxies. \newline
The lower panels represent samples where a different error cut ($30$\% on the {[N\sc ii]} and H$\alpha$ lines only) has been applied. In this way, a considerable number ($12\ 419$ from the cross-match, $398\ 272$ for the SDSS sample) of objects with weak {[O\sc iii]}, {[O\sc i]}, and H$\beta$ lines have been recovered. They appear to be mostly classified as wAGNs or RGs (as shown by the density contours). Though we miss many passive galaxies in the lower part of the upper WHAN diagrams, we found that the shape of the upper part of the distributions is almost preserved for both the radio emitters and the SDSS targets. The comparison with the lower panels suggests, as already mentioned in Sect. 2.3, that the galaxy classification is strongly influenced by the error cut we apply to the samples.\newline
Figure \ref{WHAN_oi} shows the WHAN diagnostic diagrams for the cross-matched and the SDSS samples overplotted with AGNs selected from the {[O\sc i]}-based diagram. We chose the latter as the one diagram that can classify the most Seyferts and LINERs (see Table \ref{statistics_L_ratio}). The Seyferts distribution peaks in the sAGN region of the WHAN diagram, with only few galaxies falling in the SF or wAGN regions. Moreover, it appears that most of the Seyferts do not lie close to the demarcation curve between SF and sAGN. The LINERs distributions look much more spread across the sAGN region instead, while they are supposed to be confined in the wAGN (in the case of true LINERs) or in the RG region (in case of fake LINERs). Some LINERs appear in the SF region as well. Interestingly, the latter are found to have low {[O\sc iii]}/H$\beta$, while the LINERs that fall in the RG region of the diagram have an average higher {[O\sc iii]}/H$\beta$.\newline
Figure \ref{WHAN_L} shows WHAN diagnostic diagrams for only the cross-matched sample, overplotted with Seyferts and LINERs selected from the {[O\sc i]}-based diagram. Each plot represents a different $L_{20~cm}/L_{H\alpha}$ bin (as in Figs. \ref{oi_lum_tresh}, \ref{sii_lum_tresh}, \ref{nii_lum_tresh}). While the distribution of Seyferts always peaks in the sAGN region and only the number of sources increases, LINERs are mostly present in the sAGN region in the first three bins, and they appear to be mostly located in the wAGN region only in the very last $L_{20~cm}/L_{H\alpha}$ bin. 
The radio strong (large $L_{20~cm}/L_{H\alpha}$) sources deviate from the RG and passive regions, so we can say that the radio-selection helps in removing RGs from the sample. This leads to a cleaner pure AGN sample whose LINERs may well be dominated by shocks.

\subsection{Comparison with models}
 \begin{figure*} 
 \centering
  \includegraphics[width=17.5cm]{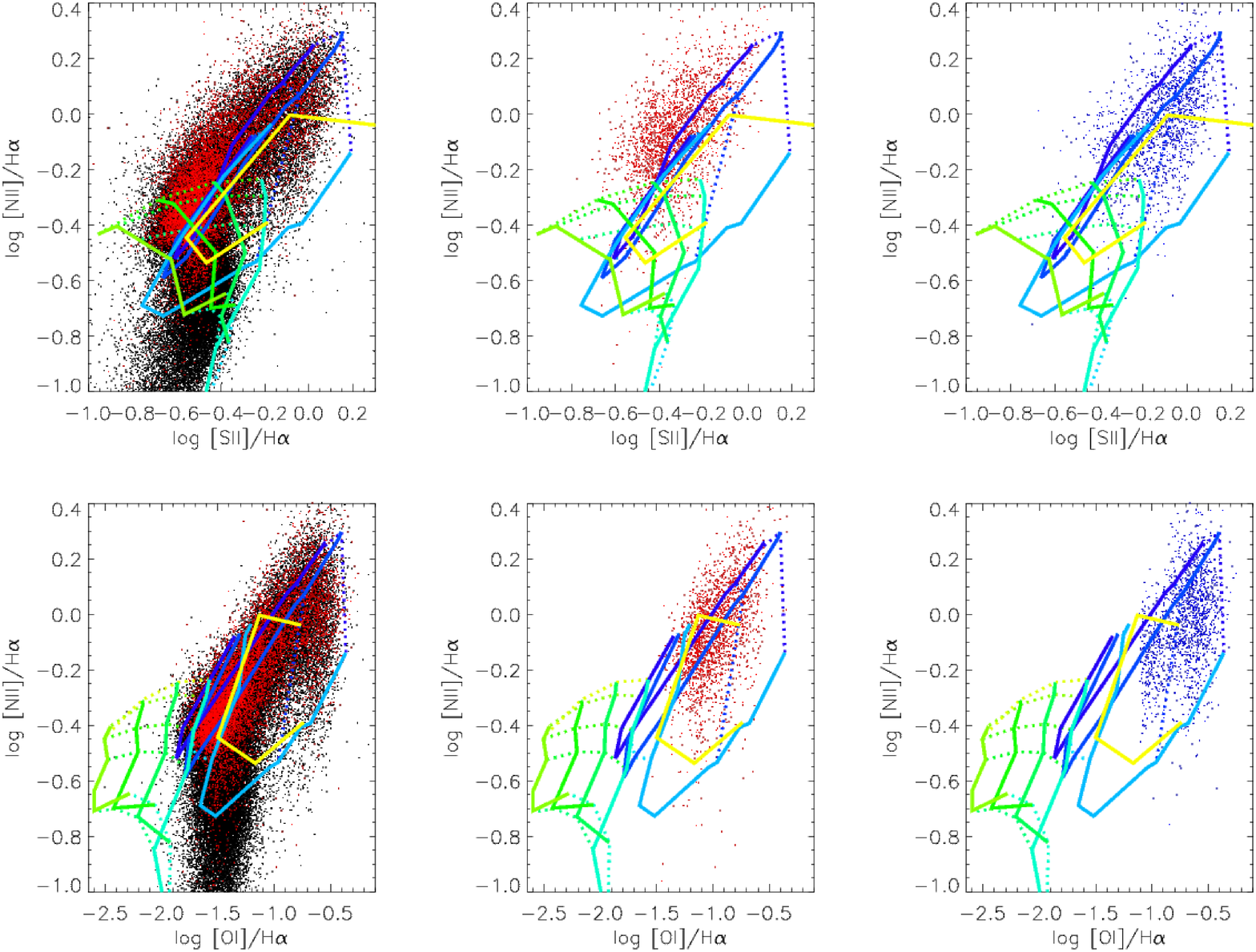}
  \caption{\label{models_comparison}Data comparison with models. Galaxies from the MPA-JHU sub-sample are in black, while the optical-radio sample is in red (left panels). In the middle panels, Seyferts of the optical-radio sample selected via {[O\sc i]}-based diagram are in red. In the right panels, LINERs selected with the same diagram are in blue. All models have $Z=2Z_{\odot}$. The star formation models \citep{Dopita2006} are represented by green lines; the dusty AGN models \citep{Groves2004} are in yellow; the models for shocks without precursors \citep{Allen2008} are in blue. In the middle panels, Seyferts of the optical-radio sample selected via {[O\sc i]}-based diagram are in red. In the right panels, LINERs selected with the same diagram are in blue.}
 \end{figure*}
By exploiting optical emission-line diagnostic diagrams, both classical and WHAN, we have stressed the higher concentration of LINERs with increasing $L_{20~cm}/L_{H\alpha}$. The point whether LINERs are pure AGNs - i.e. the main ionizing mechanism in the host galaxy is accretion onto the super-massive black hole - is still a matter of debate. There are several processes that could account for LINER emission: photoionization by a non stellar source such as an AGN accreting at a low rate \citep{Ho1993,Groves2004,Eracleous2010} and ionization by fast large-scale shocks \citep{Dopita1995} produced either by AGN radio jets, galactic super-winds \citep{Lehnert1996,Lutz1999,Cecil2001}, or starburst-driven super winds \citep{Heckman1990}. It has also been found that hot post-asymptotic giant branch (post-AGB) stars and white dwarfs can be responsible for the LINER-like emission in a considerable fraction of galaxies \citep{Binette1994}.\newline
To further explore this topic, we decided to investigate the physical mechanisms from which the observed optical emission-lines are developing. According to Dopita \& Evans (1986), the emission lines in the spectrum of individual HII regions are determined by three main physical parameters, namely the temperature of the stars, T, and thus their age, the metallicity, and the ionization parameter, U (usually defined as the ratio of the mean ionizing photon flux to the mean atom density). Using available photoionization and shocks models that can account for the line emission \citep{Groves2004,Dopita2006,Allen2008}, it is possible to unveil the dominant physical processes taking place in the galactic nuclei and their vicinity. For this purpose, we used the {[S\sc ii]}/H$\alpha$ vs. {[N\sc ii]}/H$\alpha$ and {[O\sc i]}/H$\alpha$ vs. {[N\sc ii]}/H$\alpha$ diagrams. This choice has been made by considering that these emission line ratios do not depend strongly on the age of the stellar population \citep{Moy2001,Stasinska2003,Stasinska2006}.\newline
The left-hand panels in Fig.\ref{models_comparison} show the models compared to the optical sample and the cross-matched optical-FIRST sample. We found that the position of most objects can be explained by models with $Z=2Z_{\odot}$. Solid lines represent different $\\log ~U$ values, ranging from $0$ to $-6$. Dashed lines indicate various ages, from $0.5$ Myrs to $3.0$ Myrs \citep{Dopita2006}.\newline
Models that account for AGN emission, parametrized by a smooth and featureless power law or broken power law \citep{Koski1978,Stasinska1984,Osterbrock1989,Laor1997,Prieto2000,Sazonov2004}, are mainly characterized by the ionization parameter. Dust is related to the ionization parameter by the amount of ionizing photons that can be scattered. It is thought to be a fundamental component of AGN tori, and there is also evidence of dust in the narrow-line region \citep[NLR;][]{Tomono2001,Radomski2003}. The dusty AGN model from \citet{Groves2004}, represented in yellow in Fig. \ref{models_comparison}, considers a ionization parameter ranging from $10^0$ to $10^{-4}$ and spectral index $\alpha=-2.0$. \newline
The models for shocks without precursors are from \citet{Allen2008}. The authors consider only fast shocks, where the ionizing radiation generated by the hot gas behind the shock front creates a strong photoionizing radiation field \citep{Sutherland1993,Dopita1995,Dopita1996}. The flux of the ionizing radiation emitted by the shock, hence the ionization parameter, increases with the velocity of the shock front ($\propto v^3$). The magnetic fields in the gas behind the shock front act to limit the compression through the shock itself. This effect is treated with the introduction of the magnetic parameter (${B/n_e}^{1/2}$) where the electron density ($n_e$) is equal to $1~cm^{-3}$. The different values of the magnetic parameter range from ${B/n_e}^{1/2}=0$ to $4~\mu G~cm^{3/2}$. Velocities go from $100~km~s^{-1}$ to $300~km~s^{-1}$.\newline
From the first row of plots in Fig. \ref{models_comparison} ({[S\sc ii]}/H$\alpha$ vs. {[N\sc ii]}/H$\alpha$ diagrams), it appears that we are only partially tracing the star formation in the cross-matched sample, since the left-hand side of the diagrams, where the lines indicating photoionization by hot massive stars are located, does not represent all the radio emitters. The star-forming region is indeed mostly populated by sources that do not have strong detected radio emission, so the models by \citet{Dopita2006} represent the SDSS parent sample better. The higher emission-line ratios of the rest of the radio emitters could be explained by shocks and dusty AGNs models. The lines that extend from the central part of the diagram to the upper right corner represent the full length of the distribution quite well. In the upper-middle and upper-right panels of Fig. \ref{models_comparison}, the Seyferts and the LINERs of the cross-matched optical-radio sample selected by the {[O\sc i]}-based diagram are represented. The emission-line ratios of the Seyferts cannot be fully explained by the models for shocks and dusty-AGNs that we considered in this work. On the other hand, LINERs are mostly placed in the region of the diagram that is enclosed by the sequences for fast shocks and dusty-AGNs, with only a few sources lying in the region explained by photoionization by young hot stars. The latter are found to have a rather low {[O\sc iii]}/H$\beta$ ratio. In general, LINERs shift to the upper left-hand corner of the diagram for increasing {[O\sc iii]}/H$\beta$. However, many galaxies of different spectral type are not explained by these models. The second row of Fig. \ref{models_comparison} ({[O\sc i]}/H$\alpha$ vs. {[N\sc ii]}/H$\alpha$ diagrams) shows that neither the cross-matched nor the optical samples are represented well by models for photoionization by hot massive stars. The Seyferts and LINERs are instead confined much better within the fast shocks and dusty AGNs models than in the {[S\sc ii]}/H$\alpha$ vs. {[N\sc ii]}/H$\alpha$ diagram.

\section{Discussion}
\subsection{Classical diagnostic diagrams}
With a cross-matching of optical (SDSS) and radio (FIRST) data, we obtained a sample of radio emitters with optical counterpart and thus optical spectral information. By using diagnostic diagrams, we were able to separate AGNs from starburst galaxies, studying the dependence of this classification on quantities such as $L_{20~cm}/L_{H\alpha}$ or $z$.\newline
\subsubsection{General trend of the full optical and optical-radio samples}
In our study of the optical-radio sample, we found that a relatively large number of radio emitters tend to be placed in the transitional or AGN region of the diagnostic diagrams (Fig. \ref{bpt}, Table \ref{statistics}), suggesting a strong connection between the detection of radio emission/features and the AGN activity. The percentage of detected AGNs increases from the {[S\sc ii]} ($29,4$\%) to the {[N\sc ii]} ($31,5$\%) and {[O\sc i]} ($39,1$\%) diagrams, partly due to the higher sensitivity of {[O\sc i]} to shocks. Our MPA-JHU subsample of SDSS galaxies populates a larger area of the SF region instead. The percentage of star-forming galaxies is maximum in the {[S\sc ii]} ($85,2$\%) diagram, and the percentages of pure SFs (without composites) is $67.7$\% in the {[N\sc ii]} diagram and $78.8$\% in the {[O\sc i]} diagram. This is mainly due to the selection of more starburst galaxies than in the radio-sample, where the star formation can be observed in the optical but not in the radio, below the $1$ mJy sensitivity threshold.\newline
The galaxy radio emission seems to be a function of either optical luminosity or redshift \citep{Hooper1995}. The high number of AGNs among the population of radio emitters has already been pointed out by other authors \citep[e.g. ][]{Buttiglione2010} but never extensively investigated with a set of classical and new diagnostic diagrams and by using a large sample of galaxies.
\subsubsection{Radio emitters: trend with $L_{20~cm}/L_{H\alpha}$}
We divided our sample into four equally populated bins with different ranges of log $L_{20~cm}/L_{H\alpha}$. The significant variation in the number of galaxies justifies our bin choice, although we are aware that the errors on the luminosity could affect the distribution of the sources within the bins. Although we expect different behavior for FRI and FRII, with the latter more abundant in the highest $L_{20~cm}/L_{H\alpha}$ bin, where the radio luminosity is higher, it is not possible to prove this with FIRST data alone. Owing to the large beam size of the FIRST radio survey, most of the galaxies are not structurally resolved, thus a distinction between jet-dominated and lobe-dominated emitters is not made. \newline
The optical classification displays a peak in the star-forming or composite region of the diagnostic diagrams for relatively low $L_{20~cm}/L_{H\alpha}$ values ($L_{20~cm}/L_{H\alpha}<0.716$), while the distribution appears to have a peak in the LINER region when above this threshold. We observe in all the diagrams a progressive shift of the sources towards the AGN region with increasing $L_{20~cm}/L_{H\alpha}$, indicating a change in the galaxy properties (Figs. \ref{oi_lum_tresh}, \ref{sii_lum_tresh}, and \ref{nii_lum_tresh}). We think that the observed shift in the galaxy distribution, from the SF region of the diagnostic diagrams to the Seyfert region first and LINER region could eventually be determined by two factors. First, the progressive increase in the radio luminosity peak distribution of about one order of magnitude (from $log~L_{20~cm}\sim 29.2$ to $log~L_{20~cm}\sim 30.2$) has the effect of classifying many more objects as AGNs in general and Seyferts in particular. In this respect, the larger number of AGNs suggests the radio nature of the nuclear and extended emission. Possibly, we are looking at low-luminosity AGNs producing radio-emission. Second, the different $L_{H\alpha}$ distribution of Seyferts and LINERs, with the LINERs always peaking for a particular bin at lower $L_{H\alpha}$ values with respect to Seyferts, assumes its maximum separation in the last $L_{20~cm}/L_{H\alpha}$ bin. Here $L_{H\alpha}$ of the LINERs peaks at its lowest value, $10^{41}$ erg/s, and $L_{H\alpha}$ of the Seyferts at its highest value, $10^{42}$ erg/s), thus the high $L_{20~cm}/L_{H\alpha}$ of the last bin is the reason for the appearance of a very large number of LINERs. \newline
The {[O\sc i]} diagram (Fig. \ref{oi_lum_tresh}) is the one that represents the observed increase in the number of radio emitters in the LINER region best (from $1$\% to $50$\%, see Table \ref{statistics_L_ratio}) for increasing value of $L_{20~cm}/L_{H\alpha}$. This could be due to the {[O\sc i]}/H$\alpha$ ratio, which is particularly sensitive to shocks. However, using the diagnostic diagrams might present some problems. It is well established that the activity from the AGN and the star formation inside the host are related, since they depend on the presence of fueling material (Kauffmann et al. 2003a). Galaxies that do not present strong nuclear radio activity could be detected via radio emission arising from supernovae explosions, which can contribute significantly to the emission at low radio levels. In that case, it is likely that these AGNs show up as starbursts (low tale of the radio luminosity distribution) or LINERs (high radio emission) in the classification provided by the diagnostic diagrams.
\subsubsection{Radio emitters: trend with z}
It has already been noticed in the past that there are more AGNs at high redshift and this could be related  to selection effects. At high redshift, we are likely to select powerful emitters such as Seyferts, while LINERs seem to be traced by radio-emission coming from shocks, so they are easier to find at higher radio luminosities. We explored the dependence of the optical galaxy classification on the redshift by dividing our sample of radio emitters into four different redshift bin. The classification appears to slightly depend on the redshift. The number of classified AGNs increases almost by a factor $2$ from the first to the forth redshift bins. This is true for both kinds of AGN, with a small difference between LINERs (the average factor is equal to $1.65$, based on the statistics of the {[O\sc i]} and the {[S\sc ii]} diagrams, Table \ref{statistics_L_ratio}) and Seyferts ($1.95$). This suggests that the AGN detection rate could be biased against the higher number of powerful sources, such as Seyferts, which are detected at higher redshifts. At the same time, it does not exclude AGNs and high redshift being in a genuine relation.
\subsection{Recent diagnostics}
The WHAN diagnostic diagram can help shed light on the nature of the emission lines in the LINER-like spectra of galaxies selected with the classical diagnostic diagrams. Figure \ref{WHAN_oi} shows that Seyferts (selected by using the traditional {[O\sc i]}-based diagnostic diagram) are mostly placed in the sAGN region and only a few of them are located in the wAGN region. Their number density drops at log {[N\sc ii]}/H$\alpha\lesssim -0.3$, before reaching the left boundary with the SF region. LINERs mostly appear in the wAGN region, but they are spread in the RG and sAGN regions as well. One of the possible reasons for the misclassifications of these objects could be that we selected AGNs from the {[O\sc i]}-based diagram. A full consistency between the latter and the WHAN diagram cannot be achieved.\newline
Figure \ref{WHAN_L}, as well as Figs. \ref{oi_lum_tresh}, \ref{sii_lum_tresh}, and \ref{nii_lum_tresh}, display how the separation between Seyferts and LINERs becomes more pronounced in the last $L_{20~cm}/L_{H\alpha}$ bin, where many LINERs are identified, indicating general agreement between the classification provided by the different diagnostic diagrams.\newline
We found the use of the WHAN diagram complementary to the one of the classical diagnostic diagrams, as the former is the only current optical diagnostic tool that can further distinguish between objects with LINER-like emission. In particular, the radio sources of our sample are more likely to be identified as true LINERs (wAGN) than as passive galaxies, suggesting that photoionization by old stars possibly gives only a minor contribution to line emission.
\subsection{Photoionization and shock models for the {[S\sc ii]}/H$\alpha$ vs. {[N\sc ii]}/H$\alpha$ and {[O\sc i]}/H$\alpha$ vs. {[N\sc ii]}/H$\alpha$ diagrams}
The presence of a large number of LINER-classified galaxies in the classical diagnostic diagrams does not imply a high detection rate of central engine-dominated objects. In particular, for our radio-selected sources, the observed high fraction of LINERs can be due, for example, to the detection of radio jets, as well as of SNe-driven winds. A large fraction of radio emitters are found to be AGNs (see Table \ref{statistics}), thus their emission-line ratios are expected to be at least partially explained by shock and dusty AGN models, and less by star formation. The star-forming region is, indeed, mostly populated by sources that have shallow radio emission. Models of photoionization by hot young stars \citep{Stasinska2006} can nicely explain the star-forming region (left wing of the seagull) in the classical diagnostic diagrams. Regarding the right wing of the distributions, where the AGNs are located, a comparison with photoionization models \citep{Stasinska2008} shows that radiation from old metal-rich stellar populations can explain the emission-line ratios that are typical of LINERs. On the other hand, these models do not work well for Seyfert, which could be explained by other mechanisms producing the observed emission lines.\newline
In Fig. \ref{models_comparison} ({[S\sc ii]}/H$\alpha$ vs. {[N\sc ii]}/H$\alpha$ diagram, first row), the star formation models by \citet{Dopita2006} represent the SDSS parent sample better than the cross-matched radio sample. Although the bulk of the radio emitters are placed in the region of the diagram that is enclosed by the models for star formation, shock and dusty-AGN models could account for most of the cross-matched sources that are found to be AGNs and LINERs in particular (upper-middle and upper-right panels). However, many radio emitters are placed in regions of the {[S\sc ii]}/H$\alpha$ vs. {[N\sc ii]}/H$\alpha$ diagram that the models we chose cannot predict. This is especially the case for Seyferts. This points to the possibility that these objects have emission-lines that can be explained by multiple mechanisms acting at the same time, e.g. to produce the higher {[N\sc ii]}/H$\alpha$ ratio that no model can predict. The {[O\sc i]}/H$\alpha$ vs. {[N\sc ii]}/H$\alpha$ diagram (Fig. \ref{models_comparison}, second row) does not show any Seyfert or LINER of the radio-selected sample in the region enclosed by the star formation models, but it shows more clearly that the AGNs (most of the Seyferts and almost all the LINERs) are best represented by shock and dusty AGN models. These diagrams show that, as mentioned in Sect. 3.2, the radio sources are drawn from a metal-rich population, whether starburst, LINER, or Seyfert-dominated.\newline
However, this is not sufficient to disentangle the complicated `nature of LINER´ issue. What we can say is that some LINERs are triggered by mechanisms responsible for shocks, thus they are linked to radio-emission and easily identifiable with increasing $L_{20~cm}/L_{H\alpha}$ values. We acknowledge that this evidence is just a step in unveiling the nature of LINERs, so further studies are required to provide a thorough investigation.

\section{Conclusions}
Our main findings and conclusions follow:
\begin{itemize}
 \item The requirement of detected radio emission predominantly selects active galaxies, a considerable number of AGNs and metal-rich starburst galaxies where the radio-emission mainly has a stellar origin. Radio emission correlates with optical emission-line ratios, so a cross-match allows us to classify and study radio emitters according to their optical spectral identification. 
 \item Emission-line diagnostic diagrams show that many radio emitters are classified as AGNs with increasing $L_{20~cm}/L_{H\alpha}$. In particular, the increase in $L_{20~cm}$ selects powerful radio emitters such as Seyferts, while the decrease in $L_{H\alpha}$ strongly contributes to the selection of a considerable fraction of galaxies with LINER-like emission.
 \item The higher number of identified AGNs with increasing values of $z$ can indicate a true correlation between these quantities, as well as a bias, related to the selection of the most powerful radio emitters.
 \item Emission-line diagnostic diagrams are useful tool for classifying sources according to the main ionizing mechanism producing emission lines, though the classification depends on the choice of the chemical species that is used in the diagram. In general, we found good agreement between the classifications given by the classical diagnostic diagrams and the {[N\sc ii]}/H$\alpha$ vs. equivalent width of the H$\alpha$ line (WHAN) diagram. Diagnostic diagrams can only give hints on the nature of the observed emission-lines, making a complementary comparison between data and models necessary.
 \item While the star-forming sequence in the diagnostic diagrams can be successfully fit by photoionization models, the AGN region seems to collect objects whose observed emission lines are due to different processes. Most of the radio emitters of our sample, which are mainly classified as LINERs at high  $L_{20~cm}/L_{H\alpha}$ values, have emission lines whose ratio can be explained by fast-shock and dusty-AGN models. Shocks, closely linked to the presence of radio emission, can be produced both by stars and AGNs, so that unveiling the nature of LINERs requires a more detailed study. 
 \item By using diagnostic diagrams, it is possible to select populations of LINER-like objects and further distinguish between the `true-AGNs´ and the galaxies whose emission is produced by old stars. For our radio-selected sample, LINERs are more like true AGNs than retired galaxies. Resolving the spatial radio structure of these groups of LINER-like objects, as well as Seyferts that are particularly bright in the radio, would help us shed light on the physical mechanisms that are responsible for the observed radio luminosity and optical emission-line ratios.
\end{itemize}

\begin{acknowledgements}
We are very grateful to M. Dopita for his comments and advice. We also thank the anonymous referee for the useful comments and suggestions that helped to improve the paper. M. Vitale and M. Valencia-S. are members of the International Max-Planck Research School (IMPRS) for Astronomy and Astrophysics at the Universities of Bonn and Cologne, supported by the Max Planck Society. M. Garcia-Marin is supported by the German federal department for education and research (BMBF) under project number 50OS1101. Part of this work was supported by the German Deutsche Forschungsgemeinschaft, DFG, via grant SFB 956, and by fruitful discussions with members of the European Union funded COST Action MP0905: Black Holes in a violent Universe and PECS project No. 98040. The FIRST Survey is supported in part under the auspices of the Department of Energy by Lawrence Livermore National Laboratory under contract W-7405-ENG-48 and the Institute for Geophysics and Planetary Physics. The Sloan Digital Sky Survey is a joint project of the University of Chicago, Fermilab, the Institute for Advanced Study, the Japan Participation Group, Johns Hopkins University, the Max Planck Institute for Astronomy, the Max Planck Institute for Astrophysics, New Mexico State University, Princeton University, the United States Naval Observatory, and the University of Washington. Apache Point Observatory, site of the SDSS, is operated by the Astrophysical Research Consortium. Funding for the project has been provided by the Alfred P.
Sloan Foundation, the SDSS member institutions, NASA, the NSF, the Department of Energy, the Japanese
Monbukagakusho, and the Max Planck Society. The SDSS Web site is http://www.sdss.org.
\end{acknowledgements}

\vspace*{0.5cm}
\bibliographystyle{aa} % style aa.bst
\bibliography{bib} % your references Yourfile.bib

\begin{thebibliography}{85}
\expandafter\ifx\csname natexlab\endcsname\relax\def\natexlab#1{#1}\fi

\bibitem[{{Abazajian} {et~al.}(2009){Abazajian}, {Adelman-McCarthy},
  {Ag{\"u}eros}, {Allam}, {Allende Prieto}, {An}, {Anderson}, {Anderson},
  {Annis}, {Bahcall}, \& et~al.}]{Abazajian2009}
{Abazajian}, K.~N., {Adelman-McCarthy}, J.~K., {Ag{\"u}eros}, M.~A., {et~al.}
  2009, \apjs, 182, 543

\bibitem[{{Allen} {et~al.}(2008){Allen}, {Groves}, {Dopita}, {Sutherland}, \&
  {Kewley}}]{Allen2008}
{Allen}, M.~G., {Groves}, B.~A., {Dopita}, M.~A., {Sutherland}, R.~S., \&
  {Kewley}, L.~J. 2008, \apjs, 178, 20

\bibitem[{{Baldwin} {et~al.}(1981){Baldwin}, {Phillips}, \&
  {Terlevich}}]{Baldwin1981}
{Baldwin}, J.~A., {Phillips}, M.~M., \& {Terlevich}, R. 1981, \pasp, 93, 5

\bibitem[{{Baum} \& {Heckman}(1989)}]{Baum1989}
{Baum}, S.~A. \& {Heckman}, T. 1989, \apj, 336, 702

\bibitem[{{Baum} {et~al.}(1995){Baum}, {Zirbel}, \& {O'Dea}}]{Baum1995}
{Baum}, S.~A., {Zirbel}, E.~L., \& {O'Dea}, C.~P. 1995, \apj, 451, 88

\bibitem[{{Becker} {et~al.}(1995){Becker}, {White}, \& {Helfand}}]{Becker1995}
{Becker}, R.~H., {White}, R.~L., \& {Helfand}, D.~J. 1995, \apj, 450, 559

\bibitem[{{Best} \& {Heckman}(2012)}]{Best2012}
{Best}, P.~N. \& {Heckman}, T.~M. 2012, \mnras, 421, 1569

\bibitem[{{Best} {et~al.}(2005){Best}, {Kauffmann}, {Heckman}, \&
  {Ivezi{\'c}}}]{Best2005}
{Best}, P.~N., {Kauffmann}, G., {Heckman}, T.~M., \& {Ivezi{\'c}}, {\v Z}.
  2005, \mnras, 362, 9

\bibitem[{{Binette} {et~al.}(1994){Binette}, {Magris}, {Stasi{\'n}ska}, \&
  {Bruzual}}]{Binette1994}
{Binette}, L., {Magris}, C.~G., {Stasi{\'n}ska}, G., \& {Bruzual}, A.~G. 1994,
  \aap, 292, 13

\bibitem[{{Blanton}(2006)}]{Blanton2006}
{Blanton}, M.~R. 2006, \apj, 648, 268

\bibitem[{{Blanton} {et~al.}(2003){Blanton}, {Lin}, {Lupton}, {Maley}, {Young},
  {Zehavi}, \& {Loveday}}]{Blanton2003}
{Blanton}, M.~R., {Lin}, H., {Lupton}, R.~H., {et~al.} 2003, \aj, 125, 2276

\bibitem[{{Bruzual} \& {Charlot}(2003)}]{Bruzual2003}
{Bruzual}, G. \& {Charlot}, S. 2003, \mnras, 344, 1000

\bibitem[{{Buttiglione} {et~al.}(2010){Buttiglione}, {Capetti}, {Celotti},
  {Axon}, {Chiaberge}, {Macchetto}, \& {Sparks}}]{Buttiglione2010}
{Buttiglione}, S., {Capetti}, A., {Celotti}, A., {et~al.} 2010, \aap, 509, A6

\bibitem[{{Cattaneo} {et~al.}(2007){Cattaneo}, {Blaizot}, {Weinberg}, {Kere{\v
  s}}, {Colombi}, {Dav{\'e}}, {Devriendt}, {Guiderdoni}, \&
  {Katz}}]{Cattaneo2007}
{Cattaneo}, A., {Blaizot}, J., {Weinberg}, D.~H., {et~al.} 2007, \mnras, 377,
  63

\bibitem[{{Cattaneo} {et~al.}(2009){Cattaneo}, {Faber}, {Binney}, {Dekel},
  {Kormendy}, {Mushotzky}, {Babul}, {Best}, {Br{\"u}ggen}, {Fabian}, {Frenk},
  {Khalatyan}, {Netzer}, {Mahdavi}, {Silk}, {Steinmetz}, \&
  {Wisotzki}}]{Cattaneo2009}
{Cattaneo}, A., {Faber}, S.~M., {Binney}, J., {et~al.} 2009, \nat, 460, 213

\bibitem[{{Cecil} {et~al.}(2001){Cecil}, {Bland-Hawthorn}, {Veilleux}, \&
  {Filippenko}}]{Cecil2001}
{Cecil}, G., {Bland-Hawthorn}, J., {Veilleux}, S., \& {Filippenko}, A.~V. 2001,
  \apj, 555, 338

\bibitem[{{Cid Fernandes} {et~al.}(2011){Cid Fernandes}, {Stasi{\'n}ska},
  {Mateus}, \& {Vale Asari}}]{CidFernandes2011}
{Cid Fernandes}, R., {Stasi{\'n}ska}, G., {Mateus}, A., \& {Vale Asari}, N.
  2011, \mnras, 413, 1687

\bibitem[{{Cid Fernandes} {et~al.}(2010){Cid Fernandes}, {Stasi{\'n}ska},
  {Schlickmann}, {Mateus}, {Vale Asari}, {Schoenell}, \&
  {Sodr{\'e}}}]{CidFernandes2010}
{Cid Fernandes}, R., {Stasi{\'n}ska}, G., {Schlickmann}, M.~S., {et~al.} 2010,
  \mnras, 403, 1036

\bibitem[{{Cimatti} {et~al.}(2006){Cimatti}, {Daddi}, \&
  {Renzini}}]{Cimatti2006}
{Cimatti}, A., {Daddi}, E., \& {Renzini}, A. 2006, \aap, 453, L29

\bibitem[{{Dopita} {et~al.}(2006){Dopita}, {Fischera}, {Sutherland}, {Kewley},
  {Leitherer}, {Tuffs}, {Popescu}, {van Breugel}, \& {Groves}}]{Dopita2006}
{Dopita}, M.~A., {Fischera}, J., {Sutherland}, R.~S., {et~al.} 2006, \apjs,
  167, 177

\bibitem[{{Dopita} {et~al.}(2002){Dopita}, {Groves}, {Sutherland}, {Binette},
  \& {Cecil}}]{Dopita2002}
{Dopita}, M.~A., {Groves}, B.~A., {Sutherland}, R.~S., {Binette}, L., \&
  {Cecil}, G. 2002, \apj, 572, 753

\bibitem[{{Dopita} \& {Sutherland}(1995)}]{Dopita1995}
{Dopita}, M.~A. \& {Sutherland}, R.~S. 1995, \apj, 455, 468

\bibitem[{{Dopita} \& {Sutherland}(1996)}]{Dopita1996}
{Dopita}, M.~A. \& {Sutherland}, R.~S. 1996, \apjs, 102, 161

\bibitem[{{Eracleous} {et~al.}(2010){Eracleous}, {Hwang}, \&
  {Flohic}}]{Eracleous2010}
{Eracleous}, M., {Hwang}, J.~A., \& {Flohic}, H.~M.~L.~G. 2010, \apjs, 187, 135

\bibitem[{{Fanaroff} \& {Riley}(1974)}]{Fanaroff1974}
{Fanaroff}, B.~L. \& {Riley}, J.~M. 1974, \mnras, 167, 31P

\bibitem[{{Govoni} {et~al.}(2000){Govoni}, {Falomo}, {Fasano}, \&
  {Scarpa}}]{Govoni2000}
{Govoni}, F., {Falomo}, R., {Fasano}, G., \& {Scarpa}, R. 2000, \aap, 353, 507

\bibitem[{{Groves} {et~al.}(2004{\natexlab{a}}){Groves}, {Dopita}, \&
  {Sutherland}}]{Groves2004}
{Groves}, B.~A., {Dopita}, M.~A., \& {Sutherland}, R.~S. 2004{\natexlab{a}},
  \apjs, 153, 9

\bibitem[{{Groves} {et~al.}(2004{\natexlab{b}}){Groves}, {Dopita}, \&
  {Sutherland}}]{Groves2004b}
{Groves}, B.~A., {Dopita}, M.~A., \& {Sutherland}, R.~S. 2004{\natexlab{b}},
  \apjs, 153, 75

\bibitem[{{Hardcastle} {et~al.}(2006){Hardcastle}, {Evans}, \&
  {Croston}}]{Hardcastle2006}
{Hardcastle}, M.~J., {Evans}, D.~A., \& {Croston}, J.~H. 2006, \mnras, 370,
  1893

\bibitem[{{Heckman}(1980)}]{Heckman1980}
{Heckman}, T.~M. 1980, \aap, 87, 152

\bibitem[{{Heckman} {et~al.}(1990){Heckman}, {Armus}, \& {Miley}}]{Heckman1990}
{Heckman}, T.~M., {Armus}, L., \& {Miley}, G.~K. 1990, \apjs, 74, 833

\bibitem[{{Heckman} {et~al.}(2004){Heckman}, {Kauffmann}, {Brinchmann},
  {Charlot}, {Tremonti}, \& {White}}]{Heckman2004}
{Heckman}, T.~M., {Kauffmann}, G., {Brinchmann}, J., {et~al.} 2004, \apj, 613,
  109

\bibitem[{{Ho} {et~al.}(1993){Ho}, {Filippenko}, \& {Sargent}}]{Ho1993}
{Ho}, L.~C., {Filippenko}, A.~V., \& {Sargent}, W.~L.~W. 1993, \apj, 417, 63

\bibitem[{{Ho} {et~al.}(1997){Ho}, {Filippenko}, \& {Sargent}}]{Ho1997}
{Ho}, L.~C., {Filippenko}, A.~V., \& {Sargent}, W.~L.~W. 1997, \apj, 487, 568

\bibitem[{{Hooper} {et~al.}(1995){Hooper}, {Impey}, {Foltz}, \&
  {Hewett}}]{Hooper1995}
{Hooper}, E.~J., {Impey}, C.~D., {Foltz}, C.~B., \& {Hewett}, P.~C. 1995, \apj,
  445, 62

\bibitem[{{Hopkins} {et~al.}(2006){Hopkins}, {Hernquist}, {Cox}, {Di Matteo},
  {Robertson}, \& {Springel}}]{Hopkins2006}
{Hopkins}, P.~F., {Hernquist}, L., {Cox}, T.~J., {et~al.} 2006, \apjs, 163, 1

\bibitem[{{Ivezi{\'c}} {et~al.}(2002){Ivezi{\'c}}, {Menou}, {Knapp}, {Strauss},
  {Lupton}, {Vanden Berk}, {Richards}, {Tremonti}, {Weinstein}, {Anderson},
  {Bahcall}, {Becker}, {Bernardi}, {Blanton}, {Eisenstein}, {Fan},
  {Finkbeiner}, {Finlator}, {Frieman}, {Gunn}, {Hall}, {Kim}, {Kinkhabwala},
  {Narayanan}, {Rockosi}, {Schlegel}, {Schneider}, {Strateva}, {SubbaRao},
  {Thakar}, {Voges}, {White}, {Yanny}, {Brinkmann}, {Doi}, {Fukugita},
  {Hennessy}, {Munn}, {Nichol}, \& {York}}]{Ivezic2002}
{Ivezi{\'c}}, {\v Z}., {Menou}, K., {Knapp}, G.~R., {et~al.} 2002, \aj, 124,
  2364

\bibitem[{{Kauffmann} {et~al.}(2008){Kauffmann}, {Heckman}, \&
  {Best}}]{Kauffmann2008}
{Kauffmann}, G., {Heckman}, T.~M., \& {Best}, P.~N. 2008, \mnras, 384, 953

\bibitem[{{Kauffmann} {et~al.}(2003){Kauffmann}, {Heckman}, {Tremonti},
  {Brinchmann}, {Charlot}, {White}, {Ridgway}, {Brinkmann}, {Fukugita}, {Hall},
  {Ivezi{\'c}}, {Richards}, \& {Schneider}}]{Kauffmann2003a}
{Kauffmann}, G., {Heckman}, T.~M., {Tremonti}, C., {et~al.} 2003, \mnras, 346,
  1055

\bibitem[{{Kellermann} {et~al.}(1989){Kellermann}, {Sramek}, {Schmidt},
  {Shaffer}, \& {Green}}]{Kellermann1989}
{Kellermann}, K.~I., {Sramek}, R., {Schmidt}, M., {Shaffer}, D.~B., \& {Green},
  R. 1989, \aj, 98, 1195

\bibitem[{{Kewley} {et~al.}(2001){Kewley}, {Dopita}, {Sutherland}, {Heisler},
  \& {Trevena}}]{Kewley2001}
{Kewley}, L.~J., {Dopita}, M.~A., {Sutherland}, R.~S., {Heisler}, C.~A., \&
  {Trevena}, J. 2001, \apj, 556, 121

\bibitem[{{Kewley} {et~al.}(2003){Kewley}, {Geller}, \& {Jansen}}]{Kewley2003}
{Kewley}, L.~J., {Geller}, M.~J., \& {Jansen}, R.~A. 2003, in Bulletin of the
  American Astronomical Society, Vol.~35, American Astronomical Society Meeting
  Abstracts, 119.01

\bibitem[{{Kewley} {et~al.}(2006){Kewley}, {Groves}, {Kauffmann}, \&
  {Heckman}}]{Kewley2006}
{Kewley}, L.~J., {Groves}, B., {Kauffmann}, G., \& {Heckman}, T. 2006, \mnras,
  372, 961

\bibitem[{{Koski}(1978)}]{Koski1978}
{Koski}, A.~T. 1978, \apj, 223, 56

\bibitem[{{Kozie{\l}-Wierzbowska} \& {Stasi{\'n}ska}(2011)}]{Wierzbowska2011}
{Kozie{\l}-Wierzbowska}, D. \& {Stasi{\'n}ska}, G. 2011, \mnras, 415, 1013

\bibitem[{{Lamareille} {et~al.}(2004){Lamareille}, {Mouhcine}, {Contini},
  {Lewis}, \& {Maddox}}]{Lamareille2004}
{Lamareille}, F., {Mouhcine}, M., {Contini}, T., {Lewis}, I., \& {Maddox}, S.
  2004, \mnras, 350, 396

\bibitem[{{Laor} {et~al.}(1997){Laor}, {Fiore}, {Elvis}, {Wilkes}, \&
  {McDowell}}]{Laor1997}
{Laor}, A., {Fiore}, F., {Elvis}, M., {Wilkes}, B.~J., \& {McDowell}, J.~C.
  1997, \apj, 477, 93

\bibitem[{{Ledlow} \& {Owen}(1995)}]{Ledlow1995}
{Ledlow}, M.~J. \& {Owen}, F.~N. 1995, \aj, 109, 853

\bibitem[{{Lehnert} \& {Heckman}(1996)}]{Lehnert1996}
{Lehnert}, M.~D. \& {Heckman}, T.~M. 1996, \apj, 462, 651

\bibitem[{{Lutz} {et~al.}(1999){Lutz}, {Veilleux}, \& {Genzel}}]{Lutz1999}
{Lutz}, D., {Veilleux}, S., \& {Genzel}, R. 1999, \apjl, 517, L13

\bibitem[{{Machalski} \& {Godlowski}(2000)}]{Machalski2000}
{Machalski}, J. \& {Godlowski}, W. 2000, \aap, 360, 463

\bibitem[{{McCarthy}(1993)}]{McCarthy1993}
{McCarthy}, P.~J. 1993, \araa, 31, 639

\bibitem[{{Morganti} {et~al.}(1992){Morganti}, {Ulrich}, \&
  {Tadhunter}}]{Morganti1992}
{Morganti}, R., {Ulrich}, M.-H., \& {Tadhunter}, C.~N. 1992, \mnras, 254, 546

\bibitem[{{Moustakas} {et~al.}(2006){Moustakas}, {Kennicutt}, \&
  {Tremonti}}]{Moustakas2006}
{Moustakas}, J., {Kennicutt}, Jr., R.~C., \& {Tremonti}, C.~A. 2006, \apj, 642,
  775

\bibitem[{{Moy} {et~al.}(2001){Moy}, {Rocca-Volmerange}, \& {Fioc}}]{Moy2001}
{Moy}, E., {Rocca-Volmerange}, B., \& {Fioc}, M. 2001, \aap, 365, 347

\bibitem[{{Oh} {et~al.}(2011){Oh}, {Sarzi}, {Schawinski}, \& {Yi}}]{Ossy2011}
{Oh}, K., {Sarzi}, M., {Schawinski}, K., \& {Yi}, S.~K. 2011, \apjs, 195, 13

\bibitem[{{OMullane} {et~al.}(2005){OMullane}, {Li}, {Nieto-Santisteban},
  {Szalay}, {Thakar}, \& {Gray}}]{OMullane2005}
{OMullane}, W., {Li}, N., {Nieto-Santisteban}, M., {et~al.} 2005, eprint
  arXiv:cs/0502072

\bibitem[{{Osterbrock}(1989)}]{Osterbrock1989}
{Osterbrock}, D.~E. 1989, {Astrophysics of gaseous nebulae and active galactic
  nuclei}

\bibitem[{{Pettini} \& {Pagel}(2004)}]{Pettini2004}
{Pettini}, M. \& {Pagel}, B.~E.~J. 2004, \mnras, 348, L59

\bibitem[{{Prieto} \& {Viegas}(2000)}]{Prieto2000}
{Prieto}, M.~A. \& {Viegas}, S.~M. 2000, \apj, 532, 238

\bibitem[{{Radomski} {et~al.}(2003){Radomski}, {Pi{\~n}a}, {Packham},
  {Telesco}, {De Buizer}, {Fisher}, \& {Robinson}}]{Radomski2003}
{Radomski}, J.~T., {Pi{\~n}a}, R.~K., {Packham}, C., {et~al.} 2003, \apj, 587,
  117

\bibitem[{{Rawlings} \& {Saunders}(1991)}]{Rawlings1991}
{Rawlings}, S. \& {Saunders}, R. 1991, \nat, 349, 138

\bibitem[{{Rawlings} {et~al.}(1989){Rawlings}, {Saunders}, {Eales}, \&
  {Mackay}}]{Rawlings1989}
{Rawlings}, S., {Saunders}, R., {Eales}, S.~A., \& {Mackay}, C.~D. 1989,
  \mnras, 240, 701

\bibitem[{{Sadler} {et~al.}(2002){Sadler}, {Jackson}, {Cannon}, {McIntyre},
  {Murphy}, {Bland-Hawthorn}, {Bridges}, {Cole}, {Colless}, {Collins}, {Couch},
  {Dalton}, {De Propris}, {Driver}, {Efstathiou}, {Ellis}, {Frenk},
  {Glazebrook}, {Lahav}, {Lewis}, {Lumsden}, {Maddox}, {Madgwick}, {Norberg},
  {Peacock}, {Peterson}, {Sutherland}, \& {Taylor}}]{Sadler2002}
{Sadler}, E.~M., {Jackson}, C.~A., {Cannon}, R.~D., {et~al.} 2002, \mnras, 329,
  227

\bibitem[{{Saunders} {et~al.}(1989){Saunders}, {Baldwin}, {Rawlings}, {Warner},
  \& {Miller}}]{Saunders1989}
{Saunders}, R., {Baldwin}, J.~E., {Rawlings}, S., {Warner}, P.~J., \& {Miller},
  L. 1989, \mnras, 238, 777

\bibitem[{{Sazonov} {et~al.}(2004){Sazonov}, {Ostriker}, \&
  {Sunyaev}}]{Sazonov2004}
{Sazonov}, S.~Y., {Ostriker}, J.~P., \& {Sunyaev}, R.~A. 2004, \mnras, 347, 144

\bibitem[{{Shakura} \& {Sunyaev}(1973)}]{Shakura1973}
{Shakura}, N.~I. \& {Sunyaev}, R.~A. 1973, \aap, 24, 337

\bibitem[{{Smith} \& {Heckman}(1989)}]{Smith1989}
{Smith}, E.~P. \& {Heckman}, T.~M. 1989, \apj, 341, 658

\bibitem[{{Stasi{\'n}ska}(1984)}]{Stasinska1984}
{Stasi{\'n}ska}, G. 1984, \aap, 135, 341

\bibitem[{{Stasi{\'n}ska} {et~al.}(2006){Stasi{\'n}ska}, {Cid Fernandes},
  {Mateus}, {Sodr{\'e}}, \& {Asari}}]{Stasinska2006}
{Stasi{\'n}ska}, G., {Cid Fernandes}, R., {Mateus}, A., {Sodr{\'e}}, L., \&
  {Asari}, N.~V. 2006, \mnras, 371, 972

\bibitem[{{Stasi{\'n}ska} \& {Izotov}(2003)}]{Stasinska2003}
{Stasi{\'n}ska}, G. \& {Izotov}, Y. 2003, \aap, 397, 71

\bibitem[{{Stasi{\'n}ska} {et~al.}(2008){Stasi{\'n}ska}, {Vale Asari}, {Cid
  Fernandes}, {Gomes}, {Schlickmann}, {Mateus}, {Schoenell}, {Sodr{\'e}}, \&
  {Seagal Collaboration}}]{Stasinska2008}
{Stasi{\'n}ska}, G., {Vale Asari}, N., {Cid Fernandes}, R., {et~al.} 2008,
  \mnras, 391, L29

\bibitem[{{Stoughton} {et~al.}(2002){Stoughton}, {Lupton}, {Bernardi},
  {Blanton}, {Burles}, {Castander}, {Connolly}, {Eisenstein}, {Frieman},
  {Hennessy}, {Hindsley}, {Ivezi{\'c}}, {Kent}, {Kunszt}, {Lee}, {Meiksin},
  {Munn}, {Newberg}, {Nichol}, {Nicinski}, {Pier}, {Richards}, {Richmond},
  {Schlegel}, {Smith}, {Strauss}, {SubbaRao}, {Szalay}, {Thakar}, {Tucker},
  {Vanden Berk}, {Yanny}, {Adelman}, {Anderson}, {Anderson}, {Annis},
  {Bahcall}, {Bakken}, {Bartelmann}, {Bastian}, {Bauer}, {Berman},
  {B{\"o}hringer}, {Boroski}, {Bracker}, {Briegel}, {Briggs}, {Brinkmann},
  {Brunner}, {Carey}, {Carr}, {Chen}, {Christian}, {Colestock}, {Crocker},
  {Csabai}, {Czarapata}, {Dalcanton}, {Davidsen}, {Davis}, {Dehnen},
  {Dodelson}, {Doi}, {Dombeck}, {Donahue}, {Ellman}, {Elms}, {Evans}, {Eyer},
  {Fan}, {Federwitz}, {Friedman}, {Fukugita}, {Gal}, {Gillespie}, {Glazebrook},
  {Gray}, {Grebel}, {Greenawalt}, {Greene}, {Gunn}, {de Haas}, {Haiman},
  {Haldeman}, {Hall}, {Hamabe}, {Hansen}, {Harris}, {Harris}, {Harvanek},
  {Hawley}, {Hayes}, {Heckman}, {Helmi}, {Henden}, {Hogan}, {Hogg}, {Holmgren},
  {Holtzman}, {Huang}, {Hull}, {Ichikawa}, {Ichikawa}, {Johnston}, {Kauffmann},
  {Kim}, {Kimball}, {Kinney}, {Klaene}, {Kleinman}, {Klypin}, {Knapp},
  {Korienek}, {Krolik}, {Kron}, {Krzesi{\'n}ski}, {Lamb}, {Leger},
  {Limmongkol}, {Lindenmeyer}, {Long}, {Loomis}, {Loveday}, {MacKinnon},
  {Mannery}, {Mantsch}, {Margon}, {McGehee}, {McKay}, {McLean}, {Menou},
  {Merelli}, {Mo}, {Monet}, {Nakamura}, {Narayanan}, {Nash}, {Neilsen},
  {Newman}, {Nitta}, {Odenkirchen}, {Okada}, {Okamura}, {Ostriker}, {Owen},
  {Pauls}, {Peoples}, {Peterson}, {Petravick}, {Pope}, {Pordes}, {Postman},
  {Prosapio}, {Quinn}, {Rechenmacher}, {Rivetta}, {Rix}, {Rockosi}, {Rosner},
  {Ruthmansdorfer}, {Sandford}, {Schneider}, {Scranton}, {Sekiguchi}, {Sergey},
  {Sheth}, {Shimasaku}, {Smee}, {Snedden}, {Stebbins}, {Stubbs}, {Szapudi},
  {Szkody}, {Szokoly}, {Tabachnik}, {Tsvetanov}, {Uomoto}, {Vogeley}, {Voges},
  {Waddell}, {Walterbos}, {Wang}, {Watanabe}, {Weinberg}, {White}, {White},
  {Wilhite}, {Wolfe}, {Yasuda}, {York}, {Zehavi}, \& {Zheng}}]{Stoughton2002}
{Stoughton}, C., {Lupton}, R.~H., {Bernardi}, M., {et~al.} 2002, \aj, 123, 485

\bibitem[{{Sutherland} {et~al.}(1993){Sutherland}, {Bicknell}, \&
  {Dopita}}]{Sutherland1993}
{Sutherland}, R.~S., {Bicknell}, G.~V., \& {Dopita}, M.~A. 1993, \apj, 414, 510

\bibitem[{{Tadhunter} {et~al.}(2011){Tadhunter}, {Holt}, {Gonz{\'a}lez
  Delgado}, {Rodr{\'{\i}}guez Zaur{\'{\i}}n}, {Villar-Mart{\'{\i}}n},
  {Morganti}, {Emonts}, {Ramos Almeida}, \& {Inskip}}]{Tadhunter2011}
{Tadhunter}, C., {Holt}, J., {Gonz{\'a}lez Delgado}, R., {et~al.} 2011, \mnras,
  412, 960

\bibitem[{{Tadhunter} {et~al.}(1998){Tadhunter}, {Morganti}, {Robinson},
  {Dickson}, {Villar-Martin}, \& {Fosbury}}]{Tadhunter1998}
{Tadhunter}, C.~N., {Morganti}, R., {Robinson}, A., {et~al.} 1998, \mnras, 298,
  1035

\bibitem[{{Thomas} {et~al.}(2005){Thomas}, {Maraston}, {Bender}, \& {Mendes de
  Oliveira}}]{Thomas2005}
{Thomas}, D., {Maraston}, C., {Bender}, R., \& {Mendes de Oliveira}, C. 2005,
  \apj, 621, 673

\bibitem[{{Tomono} {et~al.}(2001){Tomono}, {Doi}, {Usuda}, \&
  {Nishimura}}]{Tomono2001}
{Tomono}, D., {Doi}, Y., {Usuda}, T., \& {Nishimura}, T. 2001, \apj, 557, 637

\bibitem[{{Tortora} {et~al.}(2009){Tortora}, {Antonuccio-Delogu}, {Kaviraj},
  {Silk}, {Romeo}, \& {Becciani}}]{Tortora2009}
{Tortora}, C., {Antonuccio-Delogu}, V., {Kaviraj}, S., {et~al.} 2009, \mnras,
  396, 61

\bibitem[{{van Zee} {et~al.}(1998){van Zee}, {Salzer}, {Haynes}, {O'Donoghue},
  \& {Balonek}}]{vanZee1998}
{van Zee}, L., {Salzer}, J.~J., {Haynes}, M.~P., {O'Donoghue}, A.~A., \&
  {Balonek}, T.~J. 1998, \aj, 116, 2805

\bibitem[{{Veilleux} \& {Osterbrock}(1987)}]{Veilleux1987}
{Veilleux}, S. \& {Osterbrock}, D.~E. 1987, \apjs, 63, 295

\bibitem[{{White} {et~al.}(1997){White}, {Becker}, {Helfand}, \&
  {Gregg}}]{White1997}
{White}, R.~L., {Becker}, R.~H., {Helfand}, D.~J., \& {Gregg}, M.~D. 1997,
  \apj, 475, 479

\bibitem[{{Wild} {et~al.}(2010){Wild}, {Heckman}, \& {Charlot}}]{Wild2010}
{Wild}, V., {Heckman}, T., \& {Charlot}, S. 2010, \mnras, 405, 933

\bibitem[{{York} {et~al.}(2000){York}, {Adelman}, {Anderson}, {Anderson},
  {Annis}, {Bahcall}, {Bakken}, {Barkhouser}, {Bastian}, {Berman}, {Boroski},
  {Bracker}, {Briegel}, {Briggs}, {Brinkmann}, {Brunner}, {Burles}, {Carey},
  {Carr}, {Castander}, {Chen}, {Colestock}, {Connolly}, {Crocker}, {Csabai},
  {Czarapata}, {Davis}, {Doi}, {Dombeck}, {Eisenstein}, {Ellman}, {Elms},
  {Evans}, {Fan}, {Federwitz}, {Fiscelli}, {Friedman}, {Frieman}, {Fukugita},
  {Gillespie}, {Gunn}, {Gurbani}, {de Haas}, {Haldeman}, {Harris}, {Hayes},
  {Heckman}, {Hennessy}, {Hindsley}, {Holm}, {Holmgren}, {Huang}, {Hull},
  {Husby}, {Ichikawa}, {Ichikawa}, {Ivezi{\'c}}, {Kent}, {Kim}, {Kinney},
  {Klaene}, {Kleinman}, {Kleinman}, {Knapp}, {Korienek}, {Kron}, {Kunszt},
  {Lamb}, {Lee}, {Leger}, {Limmongkol}, {Lindenmeyer}, {Long}, {Loomis},
  {Loveday}, {Lucinio}, {Lupton}, {MacKinnon}, {Mannery}, {Mantsch}, {Margon},
  {McGehee}, {McKay}, {Meiksin}, {Merelli}, {Monet}, {Munn}, {Narayanan},
  {Nash}, {Neilsen}, {Neswold}, {Newberg}, {Nichol}, {Nicinski}, {Nonino},
  {Okada}, {Okamura}, {Ostriker}, {Owen}, {Pauls}, {Peoples}, {Peterson},
  {Petravick}, {Pier}, {Pope}, {Pordes}, {Prosapio}, {Rechenmacher}, {Quinn},
  {Richards}, {Richmond}, {Rivetta}, {Rockosi}, {Ruthmansdorfer}, {Sandford},
  {Schlegel}, {Schneider}, {Sekiguchi}, {Sergey}, {Shimasaku}, {Siegmund},
  {Smee}, {Smith}, {Snedden}, {Stone}, {Stoughton}, {Strauss}, {Stubbs},
  {SubbaRao}, {Szalay}, {Szapudi}, {Szokoly}, {Thakar}, {Tremonti}, {Tucker},
  {Uomoto}, {Vanden Berk}, {Vogeley}, {Waddell}, {Wang}, {Watanabe},
  {Weinberg}, {Yanny}, {Yasuda}, \& {SDSS Collaboration}}]{York2000}
{York}, D.~G., {Adelman}, J., {Anderson}, Jr., J.~E., {et~al.} 2000, \aj, 120,
  1579

\bibitem[{{Zirbel} \& {Baum}(1995)}]{Zirbel1995}
{Zirbel}, E.~L. \& {Baum}, S.~A. 1995, \apj, 448, 521

\end{thebibliography}

\end{document}